# Analysis of various projectile interactions with Nuclear Emulsion Detector nuclei at ~ 1 GeV per nucleon using Coulomb modified Glauber model

N. Marimuthu[1,2], V. Singh[1]*, S. S. R. Inbanathan[2]*

[1] Department of Physics, Institute of Science, Banaras Hindu University, Varanasi -221005, *INDIA*

[2] *Post Graduate and Research Department of Physics*, The *American College, Madurai-625002, INDIA*

E-mail: *venkaz@yahoo.com, stepheninbanathan@gmail.com*

Abstract

The total nuclear reaction cross-section is calculated considering with and without medium effect by employing Coulomb Modified Glauber Model (CMGM) for interactions of projectiles $^{56}Fe_{26}$, $^{84}Kr_{36}$, $^{132}Xe_{54}$, $^{197}Au_{79}$ and $^{238}U_{92}$ with nuclear emulsion detector (NED) nuclei at around 1 GeV per nucleon incident kinetic energy. These calculated reaction cross-sections are correlated with the different target groups of the NED nuclei. The average value of various parameters are also calculated and compared with the corresponding experimental results. The number of shower particles emitted in an interaction is also calculated and showed good agreement with the experimental result. We observed that the total reaction cross-section increases with increasing the target mass number in case of all the considered projectiles. In addition, it shows that the average value of reaction cross-section with nuclear medium effect is in good agreement with the experimental results for projectiles $^{56}Fe$, $^{84}Kr$, $^{132}Xe$, although results of projectiles $^{197}Au$ and $^{238}U$ are not in agreement with the experimental observations. This study sheds some light on the energy dependence of the nuclear reaction cross-section.

# 1. INTRODUCTION

The relativistic heavy ion collision in the intermediate and high-energy domains has been extensively studied both theoretically and experimentally for a long time [1–6], this study is highly interesting because of their important application and new research opportunities. In these regions, heavy-ion collision provides us information to understand the mechanism of nuclear fragmentation, space-time development of hadronic interactions under extreme condition, and formation of exotic nuclei [5-6]. The photographic nuclear emulsion detector is one of the excellent tools to understand the high-energy interactions because it provides excellent spatial resolution and very high efficiency of charge particle detection over complete solid angle [7-10]. In the heavy-ion collision, the projectile nuclei or hadrons interact with a target nucleus to produce the multi-particles and these particle productions should be considered as two different steps [5]. In the initial step, the interacting projectile nuclei mainly interacts with the primary reaction and completely overlap with the target nucleus and then leave the projectile-target participant region without any further interaction. This process associated with the production of singly charged relativistic particles i.e., shower particles ($N_S$), which is mainly pions and small mixture of k-mesons, having velocity greater than 0.7c. In the next step, the produced shower particles may be involved in the rescattering process with target nucleus and knock out the nucleons (proton) from the target nucleus. These particles are called grey particles ($N_g$). The relative velocity (v/c) of grey particle belongs in between 0.3c and 0.7c i.e. ($0.3c < \beta < 0.7c$), and their kinetic energy ranges from 30 MeV to 400 MeV i.e. ($30 < E < 400$ MeV). Due to the consequence of these two different steps, the hadron production is not an instantaneous process and it will take certain time, which is said to be creation time [5]. The excited target residuals nucleus come back to the initial state by losing their energy and attain thermal equilibrium by emitting the nuclear material in the form of fragments. These target fragments are known as black particles ($N_b$) [11-12]. These black particles have relative velocities (v/c) and kinetic energies less than 0.3c and less than 26 MeV, respectively.

The total nuclear reaction cross-section ($\sigma_R$) is one of the most important physical quantities in the heavy-ion collision. From this reaction cross-section, one can extract the fundamental information about the nuclear size and density distribution of protons and neutrons inside the nucleus [13]. Based on the nuclear reaction cross-section, one can describe the strong interaction of hadron-nucleus (h-A) and nucleus – nucleus (A-A) interactions. It has application in various research fields, including shielding against heavy-ions coming from the space radiations or accelerators, cosmic ray propagation and radio-biological effects resulting from clinical exposures [6].

The Glauber Multiple (GM) scattering theory commonly used to describe the nuclear reaction cross-section at high energies. In the high-energy collisions, the GM has been applied successfully and the total nuclear reaction cross-section has been extracted [14-16]. This model has been extended for the study of total nuclear reaction cross-section and differential elastic scattering cross-section in the lower energy domain.

In GM model, scattering amplitude defined as the phase shift function and is extended in the series, where it is describes the different multiple scattering process. The GM model is a semi-classical model picturing the nuclei moving along in the collision direction and it gives as a nucleus-nucleus collision in terms of nucleon-nucleon (NN) interaction with the given density distribution. At high energies, this model provides good approximation and in low energies, the nucleus deflected from the straight-line path due to the coulomb repulsion. This approach is called the Coulomb Modified Glauber Model (CMGM) [2, 6]. Many workers have applied CMGM in theory and experiment, and successfully calculated the total nuclear reaction cross-sections. These calculated values are found to be in good agreement with the experimental results [5, 27].

In the present work, we have calculated the total nuclear reaction cross-section for the collision of various projectiles, by using Coulomb Modified Glauber Model such as $^{56}Fe_{26}$, $^{84}Kr_{36}$, $^{132}Xe_{54}$, $^{197}Au_{79}$ and $^{238}U_{92}$ with different composition elements of the nuclear emulsion nuclei at incident energies $E_{lab} \sim 1$ GeV / n. In this model, for the reaction calculation, we consider the nuclear medium effect, because in the medium, nucleon-nucleon (NN) interactions is some cases different from the free space nucleon-nucleon interactions due to the effects of Pauli blacking and finite nuclear matter density [13]. The calculated total nuclear reaction cross-section values are compared with the corresponding projectiles experimental values. From the elements of the nuclear emulsion, we consider the two different chemical compositions according to the emulsion plates company NIKFI (BR-2), and ILFORD (G5) types.

Since, according to the simple geometrical consideration, the total number of projectile participants ($P_{proj}$), target participants ($P_{targ}$) and binary collisions ($B_c$) are calculated [19]. The participant's nucleons and binary collision involved in the collision lead to the calculation of nuclear matter effects. According to the Adamovich [22-26] empirical formula, one can easily calculate the average number of shower particles ($<N_S>$) value using the total number of participants and binary collision. Here we have calculated the average number of shower particles value and are compared with corresponding experimental results. We also studied and described the mean free path, and cross-section with the projectile mass number.

## 2. Coulomb Modified Glauber Model

According to the optical limit of the Glauber theory the total nuclear reaction cross-section for nucleus–nucleus collision can be written as [6]

$$\sigma_R (mb) = 2\pi \int [1 - T(b)] b.db. \tag{1}$$

Where, T(b) is the transparency function defined as the probability that a high energy projectile, with the impact parameter b, pass through the target without any interaction. The transparency function T(b) is calculated from the projectile and target overlap region, where interactions assumed to be single nucleon-nucleon interaction [6] and it is given by

$$T(b) = \exp[-\chi(b)]. \qquad (2)$$

Where, the imaginary part of the thickness function or nuclear phase shift function χ(b), in the case of nucleus-nucleus interaction are given by [2, 6]

$$\chi_{TP}(b) = \frac{\bar{\sigma}_{NN}}{10} \int d^2b_P \int_{-\infty}^{\infty} dz_P \int d^2b_T \int_{-\infty}^{\infty} dz_T \rho_P(b_P, z_P) \rho_T(b_T, z_T) f(b_T - (b - b_P)), \qquad (3)$$

and in the case of nucleon – nucleus interaction is written as

$$\chi_T(b) = \frac{\bar{\sigma}_{NN}}{10} \int d^2b_T \int_{-\infty}^{\infty} dz_T \rho_T(b_T, z_T) f(b_T - b). \qquad (4)$$

Where, $\bar{\sigma}_{NN}$ is the average energy dependent free space nucleon-nucleon (NN) cross-section and it is taken from the average of σ$_{nn}$ and σ$_{np}$. The ρ$_P$ and ρ$_T$ are defined as the nuclear density of the projectile and target nuclei. The function (f) is the finite range of the nucleon-nucleon interaction [2]. The nucleon–nucleon (NN) interaction cross-section at intermediate and low energies is modified with medium effect due to the Pauli blocking. The effect of Pauli blocking came from the exclusion principle and it is very essential for the internal region of inter nuclear distances owing to the high - density overlap region on the colliding nuclei. Therefore, the medium nucleon – nucleon $[(\sigma^-)_{\downarrow} NN^{\uparrow} m]$ cross-section is different from the free space nucleon-nucleon interaction cross-section $\bar{\sigma}_{NN}$ [13]. In the present work, calculation is also carryout without medium effect.

$$\bar{\sigma}_{NN} = \frac{(z_P z_T + N_P N_T)\sigma_{pp} + (z_P N_T + z_T N_P)\sigma_{np}}{A_P A_T}. \qquad (5)$$

Where, A$_P$, A$_T$, Z$_P$, Z$_T$, N$_P$ and N$_T$ are respective projectile and target mass, charge and neutron numbers. The nucleon – nucleon interaction cross-section is from the Refs. [2, 13].

$$\sigma_{nn} = \sigma_{pp} = (13.73 - 15.04\beta^{-1} + 8.76\beta^{-2} + 68.67\beta^4) \times \frac{1.0 + 7.772 E_{lab}^{0.06} \rho^{1.48}}{1.0 + 18.01 \rho^{1.46}}, \qquad (6)$$

and

$$\sigma_{np} = (-70.67 - 18.18\beta^{-1} + 25.26\beta^{-2} + 113.85\beta) \times \frac{1.0 + 20.88 E_{lab}^{0.04} \rho^{2.02}}{1.0 + 35.86 \rho^{1.90}}. \qquad (7)$$

Where, $\sigma_{PP}, \sigma_{nn}$ and $\sigma_{np}$ is the proton-proton, neutron-neutron and neutron-proton interaction cross-section and it is expressed in milli-barn (mb), $\beta = v/c$, E$_{lab}$ is the incident kinetic energy of the nucleon in MeV in the laboratory frame of reference, and ρ is the nuclear matter density in unit of fm$^{-3}$. In the equation (6) and (7), the first part describes the free space nucleon interaction and the second part describes the nuclear matter effects in the medium nucleon – nucleon interaction cross-section. The parameter β is given as [13]

$$\beta = v/c = \sqrt{1.0 - (\frac{931.5}{\frac{E_{lab}}{A_P} + 931.5})^2} \qquad (8)$$

Expression (7) used for the energy $E_{lab} > 10$ MeV and for energy, $E_{lab} < 10$ MeV, we have to use another expression as given below [2,18].

$$\sigma_{np} = [\frac{2.73}{(1-0.0553 E_n)^2 + 0.35 E_n}] + [\frac{17.63}{(1+0.344 E_n)^2 + 6.8 E_n}]. \qquad (9)$$

The projectile and target nuclear matter density distribution is assumed Gaussian in shape as given by [20, 39]

$$\rho_i(r_i) = \rho_i(0) Exp[-(b_i^2 + z_i^2)/a_i^2]. \qquad (10)$$

Where, i = (P, T); $a_i$ and $\rho_i(0)$ are the diffuseness and central nuclear density, respectively. Both of these are related to the root-mean-square radius $[(R)]_{rms}^{(i)}$, through the following expressions [20, 39]

$$\rho_i(0) = \left[\frac{A_i}{(a_i\sqrt{\pi})^3}\right] \qquad (11)$$

and

$$a_i = \sqrt{\frac{2}{3}} R_{rms}^{(i)}. \qquad (12)$$

Where i = P, T indicate projectile (P) and target (T). We used the Gaussian form function for the nucleon – nucleon range function [6].

$$f(b) = \frac{1}{\pi r_0^2} \exp(-\frac{b^2}{r_0^2}). \qquad (13)$$

Here $r_0$ parameter is related to the slope of the nucleon – nucleon differential scattering cross-section. Integrate the equation (3) with respect to the $z_P$, $z_T$, $b_T$ and $b_P$, the phase shift function χ(b), for nucleon – nucleus interaction are given as [2,5,6,21]

$$\chi_T(b) = \frac{\sqrt{\pi} \rho_T(0) a_T^3}{10(a_T^2 + r_0^2)} \overline{\sigma}_{NN} \exp(-\frac{b^2}{a_T^2 + r_0^2}). \qquad (14)$$

While the nucleus – nucleus interactions will be obtained by

$$\chi_{PT}(b) = \chi_{0PT} \exp(-\frac{b^2}{a_P^2 + a_T^2 + r_0^2}). \qquad (15)$$

Where

$$\chi_{0PT} = \frac{\pi^2 \rho_P(0) \rho_T(0) a_P^3 a_T^3}{10(a_P^2 + a_T^2 + r_0^2)} \overline{\sigma}_{NN}. \qquad (16)$$

According to the Coulomb Modified Glauber Model (CMGM), introducing the effect of Coulomb field between the projectile and target, there is a deviation in the original trajectory of the scattered particle. Therefore, the impact parameter b replaced by the b´, which relate the closest approach distance between the interacting particles [2].

$$b' = \frac{\eta + \sqrt{(\eta^2 + k^2 b^2)}}{k} \tag{17}$$

Where, k is a wave number and η is the dimensionless Sommerfeld parameter defined as

$$\eta = \frac{Z_P Z_T e^2}{\hbar v}. \tag{18}$$

Where, $Z_P e$, $Z_T e$ are the total charge of the projectile and target nucleus, respectively and v is the velocity of the projectile in unit of c. It should be mentioned that, in all our calculations, the overlap integral of Eqns. (3) and (4) are evaluated in terms of the b´. On substituting the Eqn. (15) into Eqn. (2), one can calculate the total nuclear reaction cross-section ($\sigma_R$) for the proton and for the different projectiles interactions with different targets i.e. constituents of nuclear emulsion detector. These calculated nuclear reaction cross-sections used in the calculation of average number of projectile participants [(P)$_{proj}$], target participants [(P)$_{targ}$] and binary collision [(B)$_C$] through the following simple geometrical consideration [19].

$$<P_{proj}> = \frac{A_P \sigma_{PA_T}}{\sigma_{A_P A_T}}, \tag{19}$$

$$<P_{targ}> = \frac{A_P \sigma_{PA_P}}{\sigma_{A_P A_T}}, and \tag{20}$$

$$<B_C> = \frac{A_P A_T \sigma_{nn}}{\sigma_{A_P A_T}}. \tag{21}$$

In the above equations, $\sigma_{PA_T}$ is the total nuclear reaction cross-section of the proton with target i.e., each target belongs to the nuclear emulsion detector constituent, consider as a target $\sigma_{PA_P}$ and $\sigma_{nn}$ are the total nuclear reaction cross-section of the proton with projectile, and proton-proton cross-section. In addition, $\sigma_{A_P A_T}$ is the total nuclear reaction cross-section of the projectile. The average numbers of projectile participants [(P)$_{proj}$], target participants [(P)$_{targ}$] and binary collision [(B)$_C$] is used in the shower particle multiplicity calculation.

## 3. Results and Discussions

We have used the approach discussed in the sec. 2 for the calculation of total nuclear reaction cross-section for proton-Emulsion (p-Em), $^{56}$Fe-Em, $^{84}$Kr-Em, $^{132}$Xe-Em, $^{197}$Au-Em, and $^{238}$U-Em at incident engines ~1 GeV per nucleon. These calculations have been performed in the Coulomb modified Glauber model (CMGM) environment using parameters related to the free space nucleon-nucleon interaction ($\bar{\sigma}_{NN}$) and medium nucleon-nucleon interaction $[(\sigma^-)_{NN}^m]$. Consideration taken in the calculation of total nuclear reaction cross-section ($\sigma_R$) is ρ = 0, in case of without nuclear medium effect. In case of with nuclear medium effect, we used ρ = 0.15 fm$^{-3}$, ρ = 0.17 fm$^{-3}$ and ρ = 0.19 fm$^{-3}$, for the calculation of total nuclear reaction cross-section. Here, ρ is the saturation density of the normal nuclear matter, which ranges from 0.15–0.19 fm$^{-3}$ [13]. The calculated nuclear reaction cross-section with medium effects is represented as $\sigma_{R1}^m, \sigma_{R2}^m$, and $\sigma_{R3}^m$. This NN interaction used in this calculation defined as the medium nucleon - nucleon interaction $[(\sigma^-)_{NN}^m]$ and generally most of the previous calculations [27, 28], considered nuclear matter density ρ = 0.17 fm$^{-3}$ only, in their $\bar{\sigma}_{NN}^m$ calculation. It is worth mentioning here that, we have performed all theoretical calculation of nuclear reaction cross-section in accordance with the zero - range approach. These nuclear reaction cross-section values are plotted with respect to the mass number of the different target of the nuclear emulsion detector nuclei for different projectiles at incident kinetic energy around 1 GeV per nucleon in figures 1 and 2.

From figures 1 and 2, we may see that the total nuclear reaction cross-section with and without nuclear medium effect is increases with the mass number of target nucleus, in case of all considered projectiles. From these figures 1 & 2, one can also observe that the nuclear reaction cross-section $\sigma_R, \sigma_{R1}^m, \sigma_{R2}^m, and \sigma_{R3}^m$ have very close value to each other in case of the target mass number $A_T < 40$. The total nuclear reaction cross-section with medium effect, $\sigma_{R1}^m, \sigma_{R2}^m, and \sigma_{R3}^m$ have no significant dependence in the mentioned range of the nuclear matter density (ρ). As shown in the figure 1, there are many calculated / theoretical values of nuclear reaction cross-section for proton – emulsion, however the nuclear reaction cross-section without medium effect [(σ)$_R$] have always higher values than the nuclear reaction cross-section with medium effects. Figure 2, shows that the total nuclear reaction cross-section of the heavy projectiles - emulsion interactions for different projectiles as a function of the target mass. The nuclear reaction cross-section, $\sigma_R, \sigma_{R1}^m, \sigma_{R2}^m, and \sigma_{R3}^m$ values for $^{56}$Fe-Em and $^{84}$Kr-Em values are close to each other, however the medium effect values for $\sigma_{R1}^m, \sigma_{R2}^m, and \sigma_{R3}^m$ are dominating in case of $^{132}$Xe - Em and $^{238}$U - Em, and some place shows protrusions and it may be due to the variation in the projectile mass and nuclear matter density. The nuclear reaction cross-section of $^{197}$Au-Em shows the clear difference of medium effect (σ)$_R$ and without medium effect and in that graph, the medium effect nuclear reaction cross-section (σ)$_R$ values higher than other three values $\sigma_{R1}^m, \sigma_{R2}^m, and \sigma_{R3}^m$.

The calculated total nuclear reaction cross-section with medium and without medium, in case of proton-Emulsion and $^{56}$Fe - Em, $^{84}$Kr - Em, $^{132}$Xe - Em, $^{197}$Au - Em, and $^{238}$U - Em at ~ 1 GeV/n are tabulated in Tables 1 and 2. It is important to note that, the calculated nuclear reaction cross-section with medium effect $\sigma_{R2}^m$ is only given in the Table because a previous graph shows, $\sigma_{R1}^m, \sigma_{R2}^m, and \sigma_{R3}^m$ values are more close to each other. It should be mentioning that; the first column of the Table refers to the constituent elements of the nuclear emulsion, while the second and third column related to the mass number and root-mean-square (rms) radius of the corresponding elements. In the present study, the rms charge radius is not available for the three projectile nuclei, $^{56}$Fe, $^{132}$Xe and $^{197}$Au and it has been calculated using the following global expression of Ref. [29]

$$R_{rms} = 0.891 A^{-\frac{1}{3}}(1+1.565 A^{-\frac{2}{3}} - 1.04 A^{-\frac{4}{3}}). \tag{22}$$

The fifth column referred to the chemical concentration of the element nuclei, NIKFI (Br-2) type of emulsion used for $^{56}$Fe - Em, $^{84}$Kr - Em and ILFORD (G5) type emulsion used in the case of $^{132}$Xe - Em, $^{197}$Au - Em, $^{238}$U - Em. The sixth and seventh column referred that the calculated total nuclear reaction cross-section without medium ($\sigma_R$) and with medium effect ($\sigma_{R2}^{m}$) for interaction of proton - Emulsion and different projectiles with the emulsion. The multiplied total nuclear reaction cross-section $\sigma_R^{(ml)}$ and $\sigma_R^{m(ml)}$ were obtained from the calculated total nuclear reaction cross-section $\sigma_R$ and $\sigma_{R2}^{m}$ for the different projectiles multiplied with individual emulsion nuclei's ($^{1}$H, $^{12}$C, $^{14}$N, $^{16}$O, $^{32}$S, $^{80}$Br, $^{108}$Ag, $^{127}$I) chemical concentration. The summation value of the $\sigma_R^{(ml)}$ and $\sigma_R^{m(ml)}$ is divided by the sum of the chemical concentration of the elements of the emulsion, one can get the average value of the total reaction cross-section of the proton – emulsion and different projectiles with emulsion ($^{56}$Fe - Em, $^{84}$Kr - Em, $^{131}$Xe - Em, $^{197}$Au - Em, $^{238}$U - Em) for the complete sample. This calculated average value of the total nuclear reaction cross-section has been compared with corresponding experimental data. It should be noted that, the experimental total nuclear reaction cross-section has been obtained from the following expression [33]

$$\sigma_R^{\exp} = \frac{1}{n_{cc} \lambda_{\exp}}. \tag{23}$$

Where, $\lambda_{\exp}$ is the mean free path of the experimental value and $n_{cc}$ is the summation value of the chemical concentration of the elements of emulsion. The experimental mean free path value is playing an important role in the calculation of total nuclear reaction cross-section.

In the figure 3, we have plotted the interaction mean free path of different projectiles ($^{16}$O, $^{40}$Ar, $^{12}$C, $^{14}$C, $^{24}$Mg, $^{28}$Si, $^{32}$S, $^{7}$Li, $^{56}$Fe, $^{84}$Kr, $^{132}$Xe, $^{197}$Au, $^{238}$U) as a function of the projectile mass number. It is important to note that; here all plotted projectiles have different incident energy except for projectiles $^{56}$Fe; $^{84}$Kr; $^{132}$Xe, $^{197}$Au, $^{238}$U, and the mean free path taken for these projectiles are from Refs. [5, 35 - 41]. In figure 3, the (top) first graph includes the mean free path (mfp) of $^{197}$Au projectile and the (bottom) second one excluded the mfp of $^{197}$Au projectile, because the $^{197}$Au projectile is only showing the different anomalous effect compared with other projectiles. It is evident from figure 3 that the mean free path gradually decreases with increasing the projectile mass number. These results indicate that the mean free path strongly depends on the projectile mass number and have weak dependency on the projectile energy.

The calculated projectile-emulsion average value of total nuclear reaction cross-section considered with and without nuclear medium effect compared with the corresponding experimental results for different projectiles such as $^{56}Fe_{26}$, $^{84}Kr_{36}$, $^{132}Xe_{54}$, $^{197}Au_{79}$ and $^{238}U_{92}$ at ~1 GeV/n are shown in the figure 4 (a & b). In figure 4 (a & b), the solid circles represent the experimental reaction cross-section for the above-mentioned projectiles. The calculated average value of the nuclear reaction cross-section without considering the nuclear medium effect is presented in figure 4(a) and it is represented as $\sigma_R^{theory}$. These calculations have been done using nuclear matter density $\rho = 0$. The average value of nuclear reaction cross-section considers with nuclear medium effect is displayed in the figure 4(b). These calculations have been done considering with nuclear matter density $\rho = 0.15$ fm$^{-3}$, $\rho = 0.17$ fm$^{-3}$ and $\rho = 0.19$ fm$^{-3}$ and it is represented as the $\sigma_{R1}^{(m)theory}, \sigma_{R2}^{(m)theory} and \sigma_{R3}^{(m)theory}$. As shown in the figure 4 (a & b), the average value of nuclear reaction cross-section continually increases with increasing projectile mass and the calculated theoretical values are always higher than the experimental one. From figure 4(a), one can observe that the calculated value $\sigma_R^{theory}$ shows reasonable agreement with corresponding experimental values for projectiles $^{56}$Fe-Em, $^{84}$Kr-Em and $^{132}$Xe-Em, and it shows disagreement for the projectiles $^{197}$Au-Em and $^{238}$U-Em. From figure 4(b), the calculated nuclear reaction cross-section $\sigma_{R1}^{(m)theory}, \sigma_{R2}^{(m)theory}$ and $\sigma_{R3}^{(m)theory}$ are more close to each other, and from this figure the average value of nuclear reaction cross-section increases with increasing nuclear matter density. The experimental values are in good agreement with the calculated ones for the projectiles $^{56}$Fe-Em, $^{84}$Kr-Em and $^{132}$Xe-Em, and disagreement for the projectiles $^{197}$Au-Em and $^{238}$U-Em.

From these graphs [figures 4 (a & b)], one can observe that the $^{197}$Au projectile experimental nuclear reaction cross- section and predicted nuclear reaction cross-section show large difference. It is due to the experimental nuclear reaction cross-section value that highly suppressed for the $^{197}$Au projectile. The reason behind this is unknown. The experimental nuclear reaction cross-section value has been calculated using the mean free path ($\lambda$) value of $^{197}$Au projectile [34]. From figure 4b, we can conclude that the calculated nuclear reaction cross-section with medium effect shows good agreement with experimental values within the statistical error except for $^{197}$Au. The $^{197}$Au projectile only shows the anomalous effect compared to the other projectiles, so it is very important to recheck the $^{197}$Au projectile experimental mean free path value. From the above results, we may conclude that introducing the nuclear medium effect is necessary for the Coulomb modified Glauber model for the descriptions of heavy-ion collision. We may also conclude that the nuclear reaction cross-sections with medium effect have lower value than the without medium effect. These changes may occur by decreases of nucleon-nucleon (NN) interaction cross-section due to the consideration of nuclear medium. The calculation of nuclear reaction cross-section not only depends on the radii of projectile and target but also it depends on the projectile mass and medium.

In figure 5, the energy dependence of the average value of the total nuclear reaction cross-sections are shown. It shows that the calculated average value of the total nuclear reaction cross-section [$\sigma_R^{theory}$] without nuclear medium effect for projectiles $^{56}Fe_{26}$, $^{84}Kr_{36}$, $^{132}Xe_{54}$, $^{197}Au_{79}$ and $^{238}U_{92}$ at ~1 GeV / n are compared with $^{16}O_{32}$ projectile of different energy from 0.2 GeV to 200 GeV. The calculated and experimental values for $^{16}O_{32}$ projectile are taken from Ref. [5]. It can be seen from the figure 5 that the calculated nuclear reaction cross-section for $^{16}O$-Em at 0.2 GeV/n and $^{56}Fe$-Em, $^{84}Kr$-Em and $^{132}Xe$-Em at ~1 GeV/n are showing reasonable agreement with the experimental values. The $^{16}O$-Em interactions above 2 GeV show significant disagreement with the experimental results and it also shows disagreement with the higher mass projectiles such as $^{197}Au$-Em and $^{238}U$-Em. It reflects that the present model is not suitable for the higher-mass and higher-energy projectiles and further modification should be considered.

In the figure 6, we displayed the ratio of $\sigma_{Expt}/\sigma_{Cal}$ as the function of projectile mass for $^{56}Fe_{26}$, $^{84}Kr_{36}$, $^{132}Xe_{54}$, $^{197}Au_{79}$ and $^{238}U_{92}$ at ~1 GeV/n. From figure 6, one can see that the model predicted value of the nuclear reaction cross-section is close to the experimentally measured nuclear reaction cross-section value for the projectiles $^{56}Fe_{26}$, $^{84}Kr_{36}$ and $^{132}Xe_{54}$. However, it fails to predict the same for the projectiles $^{197}Au_{79}$ and $^{238}U_{92}$. It shows that the proposed model needs further modification to explain the similar phenomena for heavy nuclei such as $^{197}Au_{79}$ and $^{238}U_{92}$.

Using the CMGM approaches, we have also calculated the average number of projectile participants ($P_{proj}$), target participants ($P_{targ}$) and binary collision ($B_C$) over the different constituents of nuclear emulsion detector for the interaction of different projectiles. The obtained values are tabulated in Table 3 & Table 4. For the calculation of above mentioned parameters, we used the Eqns. (19), (20) and (21). In these equations, nucleus-nucleus $\sigma_{(A_P A_T)}$, proton-nucleus $\sigma_{(P_P P_T)}$ and proton-proton $\sigma_{(PP)}$ cross-section is parameterized as [33],

$$\sigma_{(A_P A_T)}(mb) = 109.2\ (A_P^{0.29} + A_T^{0.29} - 1.39)^2, \tag{24}$$

$$\sigma_{(P_P P_T)}(mb) = 38.17\ A^{0.719} \tag{25}$$

$$\sigma_{(PP)}(mb) = 32.3(mb) \tag{26}$$

From Table 3 & 4, one can see that, the values of projectile participants, target participants and binary collision for the interactions of any considered projectiles with the corresponding nuclear emulsion detector's target nuclei increases as the mass number of the target nuclei is increasing. From the table 3, one can observe that, the projectile participant has unit value for the interaction of proton with the nuclear emulsion detector's nucleus except in the case of hydrogen target nucleus. On the other hand, the value of projectile participants and target participants, for the interaction of proton with H-nucleus is same for the projectiles $^{56}Fe_{26}$, $^{84}Kr_{36}$, $^{132}Xe_{54}$, $^{197}Au_{79}$ and $^{238}U_{92}$. In the case of proton-Em, the summed values of $P_{proj}$ and $P_{targ}$ in the reactions $^{56}Fe$-$^{1}H$, $^{84}Kr$-$^{1}H$, $^{132}Xe$-$^{1}H$, $^{197}Au$-$^{12}C$, $^{238}U$-$^{1}H$ or $^{56}Fe$-$^{12}C$, $Kr$-$^{12}C$, $^{132}Xe$-$^{12}C$, $^{197}Au$-$^{12}C$, $^{238}U$-$^{12}C$ etc. is usually higher than the values of the binary collisions. The values of target participants and binary collision, in the case of nuclear medium effects and proton – proton collision for the interaction of various projectiles with the same target nucleus of the nuclear emulsion detector (such as $^{56}Fe$-$^{12}C$, $^{84}Kr$-$^{12}C$, $^{132}Xe$-$^{12}C$, $^{197}Au$-$^{12}C$ and $^{238}U$-$^{12}C$ etc.), decreases with increase in the mass number of the projectiles. From the Table 4, we may see that the same parameters are increasing with the increase in the projectile mass number in case of nucleus – nucleus collisions and keeping rest conditions same as above i.e. in case of Table 3. From the Table 4, it is clear that the calculated total nuclear reaction cross-section with and without nuclear medium effects are larger in case of nucleus – nucleus collisions than the values obtained in case of proton – proton collision, which are tabulated in the Table 3. This may be due the multiple number of collisions among the nucleons of the two colliding nucleus. From Tables 3 and 4, we may conclude that the number of projectile participants, target participants and the participant from the binary collision are heavily dependent on the mass number of the colliding nuclei that also strongly supports the theory of superposition of nucleon i.e. multiple collisions during nucleus – nucleus interactions.

The calculated number of average projectile participants and target participants, and average number of binary collision for different projectile have been tabulated in Table 4. Considering the CMGM model approach and without nuclear matter effects, the average number of participants i.e. the sum of projectile and target participants is 27.39 and 37.56 number of binary collision are obtained for the case of $^{84}Kr_{36}$ – Em interactions at around 1 GeV per nucleon kinetic energy. In the case of proton – proton collision, the same parameters for $^{84}Kr_{36}$ – Em interactions are 3.56 and 2.10, respectively. The ratio of nucleus – nucleus and proton – proton collisions for the number of participating nucleons in the collision and the number of binary interactions in a collision will shed some light in the growth of the amount of nuclear matter involved in the collision of proton – proton to nucleus – nucleus, and the rations for the $^{84}Kr_{36}$ – Em interactions are 7.69 and 17.88, respectively. The estimate of average number of produced pions and kaons in an interaction can be obtained using following equations [36].

For the proton – nucleus/proton collisions:

$$<n_s>_{P-Em} = 2.34 <n_s>_{PP} - 4.12. \qquad (27)$$

From the nucleus – nucleus collisions:

$$<n_s>_{^{84}Kr_{36}-Em} = 17.99 <n_s>_{PP} - 31.68. \qquad (28)$$

And for the binary collision approach:

$$<n_s>_{^{84}Kr_{36}-Em} = 41.83 <n_s>_{PP} - 73.66. \tag{29}$$

From the equation (27), one can understand that the average multiplicity of singly charged relativistic particles in the proton – emulsion interactions ($<n_s>_{P-Em}$), linearly depends on the charged particle multiplicity in the case of proton – proton interaction ($<n_s>_{PP}$), at the same energy [22-26]. Addition of the resultant values of equation (28) and (29), one can get the average multiplicity of singly charged relativistic particles for $^{84}$Kr-Em interaction at ~1GeV/n. Following the above mentioned procedures, we obtained the value of charged relativistic particles for different reaction $^{56}$Fe-Em, $^{132}$Xe-Em, $^{197}$Au-Em and $^{238}$U-Em. It is worth mentioning here that in a similar fashion we have calculated the value of the relativistic charged particles for medium effect. The calculated average value of the singly charged relativistic particles or shower particle's multiplicities are compared with the corresponding experimental values are shown in the table 5.

From the table 5, we may see that the calculated shower particles (i.e. produced particles during collisions) multiplicities from the Coulomb Modified Glauber Model (CGCM) consideration with and without medium effect successfully reproduce the experimental results within statistical error. Since, the calculated participant multiplicity values are less than the binary collision values according to the CGCM calculation. Therefore, it indicates that the expected large number of shower particles is mainly coming from the binary collision in case of all reactions. This effect may be seen for both cases with and without nuclear medium effect. We can also observe that the experimental and calculated values of shower particles multiplicities are linearly increasing with increase in the projectiles mass number and incident kinetic energy [5].

## 4. Conclusions

In this work, we have calculated and compared the total nuclear reaction cross-section for proton – emulsion and nucleus – emulsion interactions considering with and without nuclear medium effects for large number of projectiles such as $^{56}$Fe$_{26}$, $^{84}$Kr$_{36}$, $^{132}$Xe$_{54}$, $^{197}$Au$_{79}$ and $^{238}$U$_{92}$ at the incident kinetic energy of ~ 1 GeV per nucleon in the framework of Coulomb Modified Glauber Model (CMGM). For the calculation of nuclear reaction cross-section with nuclear medium effect, we considered different values of nuclear matter density. All theoretical calculations presented in this paper are performed in zero range approach. The small change in nuclear matter density of nucleon-nucleon (N-N) interaction cross-section does not produced any remarkable changes in nucleus-nucleus (A-A) interactions cross-section and also in the proton – emulsion interactions cross-section. However, comparison results represent that introducing the nuclear medium effect is necessary for the Coulomb Modified Glauber Model for the descriptions of heavy-ion collision. The total nuclear reaction cross-sections are also calculated for the different constituents of the nuclear emulsion detector's nuclei. From the results, it may be concluded that the all obtained total nuclear reaction cross-sections increases with increase in the target mass number. In most of the cases, the average values of projectile - emulsion nuclear reaction cross-section are compared with the corresponding experimental results and shows significant agreements. We observed that the nuclear reaction cross-sections with nuclear medium effect predicts less value than the without nuclear medium effect. These changes may be possible due to decrease in the nucleon – nucleon (N-N) cross-section in the presence of nuclear medium. Since, the average value of nuclear reaction cross-section is continuously increases with increase in the projectile mass number. However, we observed that the calculated results without nuclear medium effect are in fairly good agreement with experimental results of projectiles $^{56}$Fe, $^{84}$Kr, $^{132}$Xe and shows disagreement with results of projectiles $^{197}$Au and $^{238}$U. We observed similar results in case of nuclear medium effect. Since, results obtained from CMGM analysis for $^{197}$Au projectile only differs strongly with the experimental results. Therefore, it is important to pay more attention on the experimental result of the $^{197}$Au projectile mean free path. The nuclear reaction cross-section is not only depends on the radii of the projectile and target but also depends on the projectile mass number and the nuclear medium present. The calculated average value of nuclear reaction cross-section without nuclear medium effect has been compared with different energy regions of $^{16}$O projectile and showed reasonably good agreement with calculated results. It may be an indication that the model should be modified for high mass and energy regions. The experimental mean free path value has strong influence on the experimental nuclear reaction cross-section. We observed the same with weak dependence of kinetic energy of the projectile. The ratio of $\sigma_{Expt}/\sigma_{Cal}$ as the function of projectile mass graph revealed that the CMGM prediction for the nuclear reaction cross-sections are very close to the experimental results for the projectiles $^{56}$Fe, $^{84}$Kr and $^{132}$Xe. We have also calculated the number of binary collisions for the proton-emulsion and nucleus-emulsion interactions cross-sections considering with and without nuclear medium effect. These average values are used in calculation of the average values of the shower particles. The calculated values of shower particles for different projectiles are compared with the corresponding experimental value and found in good agreement. The shower particles multiplicities depend on the projectile mass and as well as incident energy of the projectiles.

## 5. Acknowledgement

Authors are thankful to the Department of Science and Technology (DST), New Delhi for their financial support.

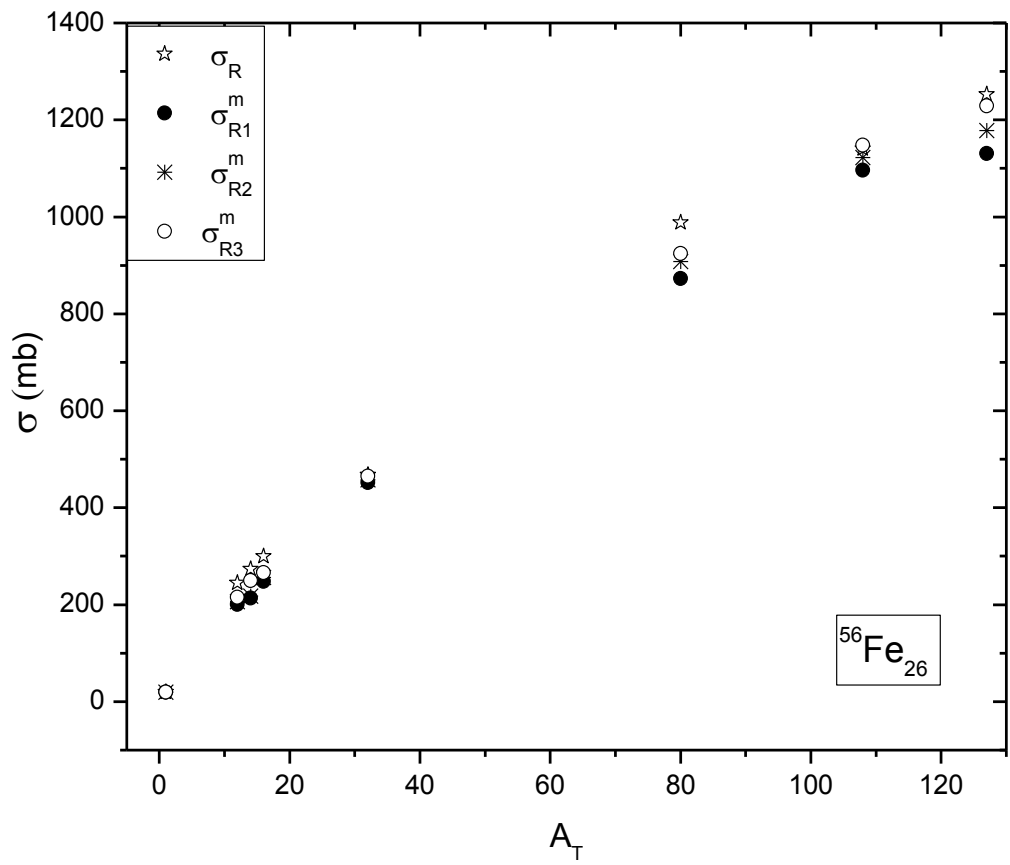

**(a)**

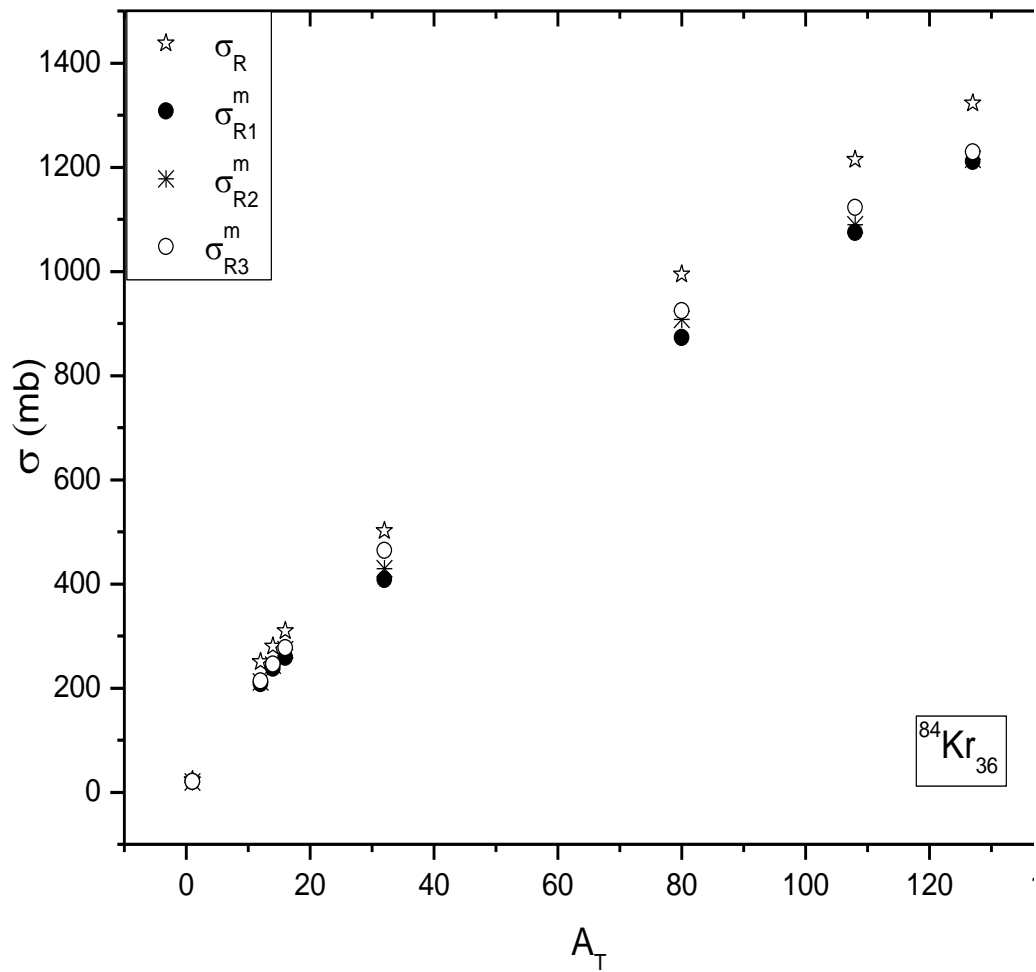

**(b)**

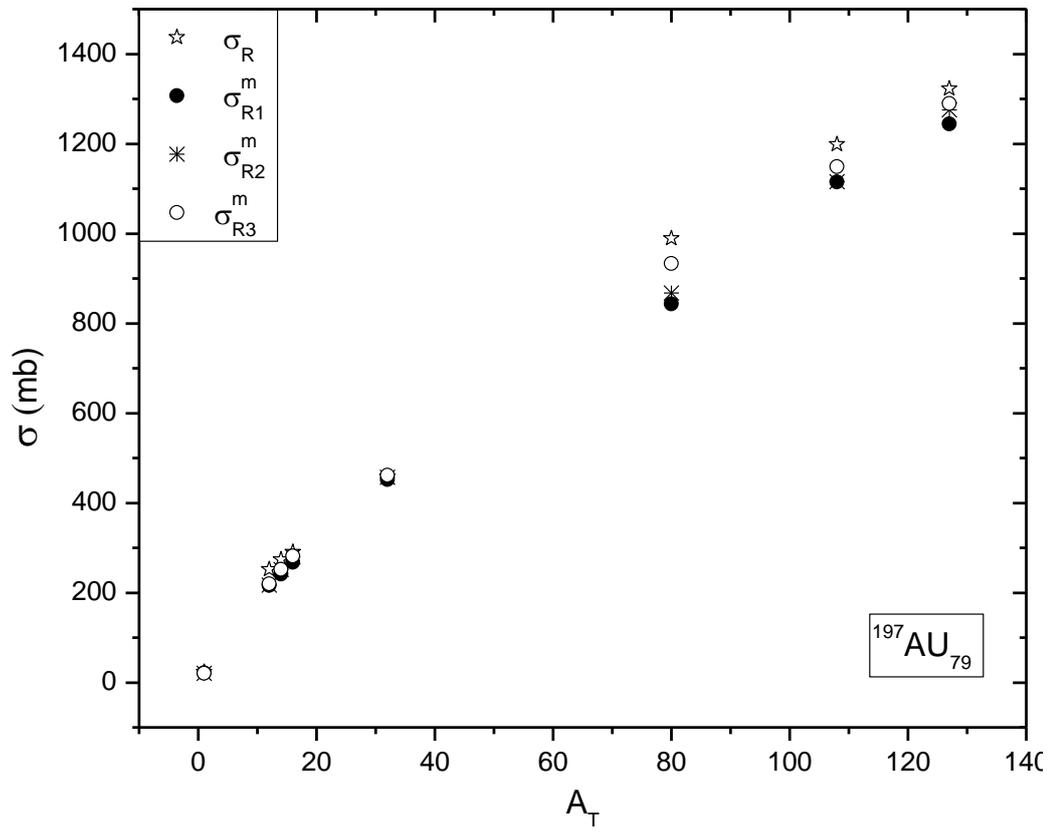

**(c)**

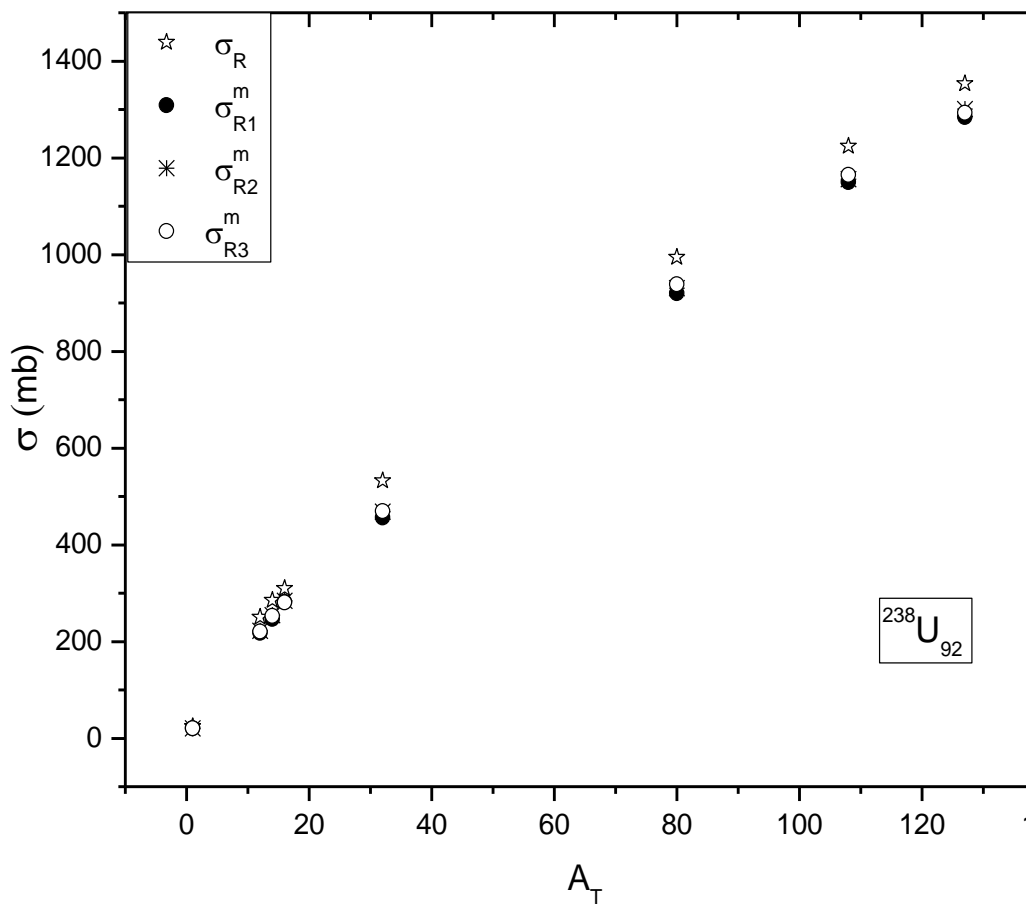

**(d)**

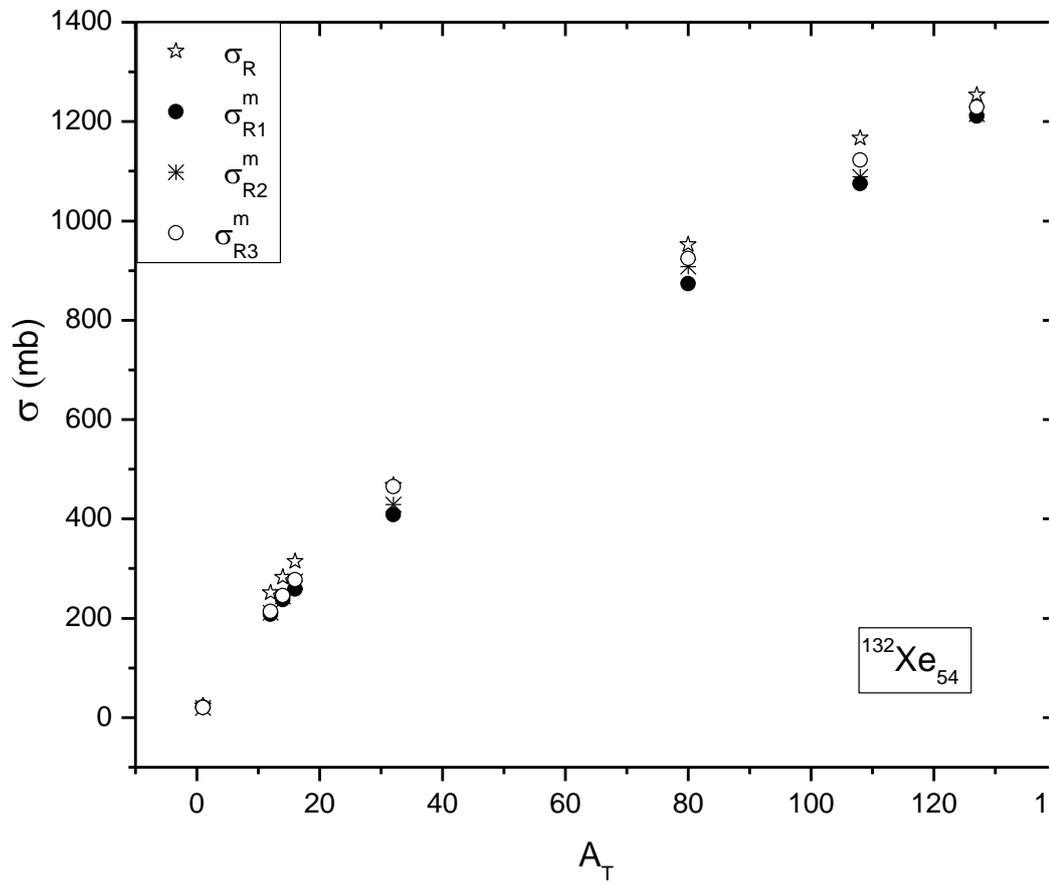

**(e)**

**Figure 1:** The total nuclear reactions cross-section of the proton - emulsion with medium and without medium effect for the different projectiles as function of emulsion target mass ($A_T$) are shown from (a) to (e).

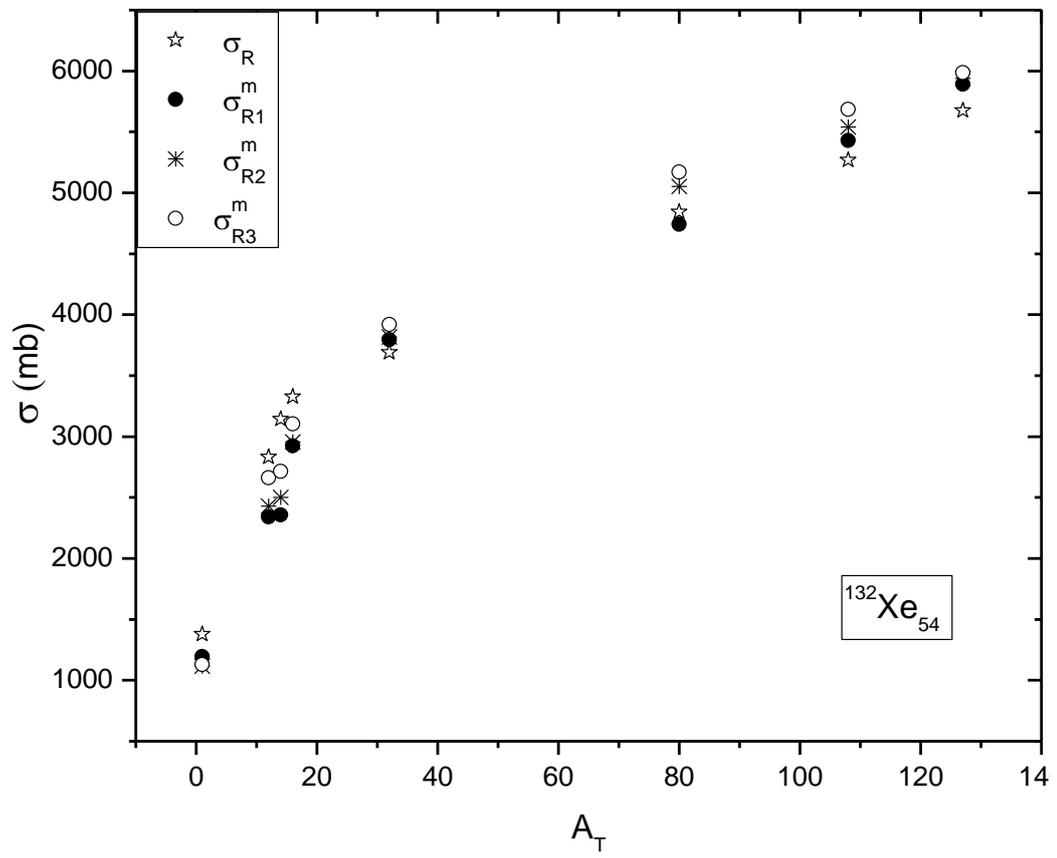

**(a)**

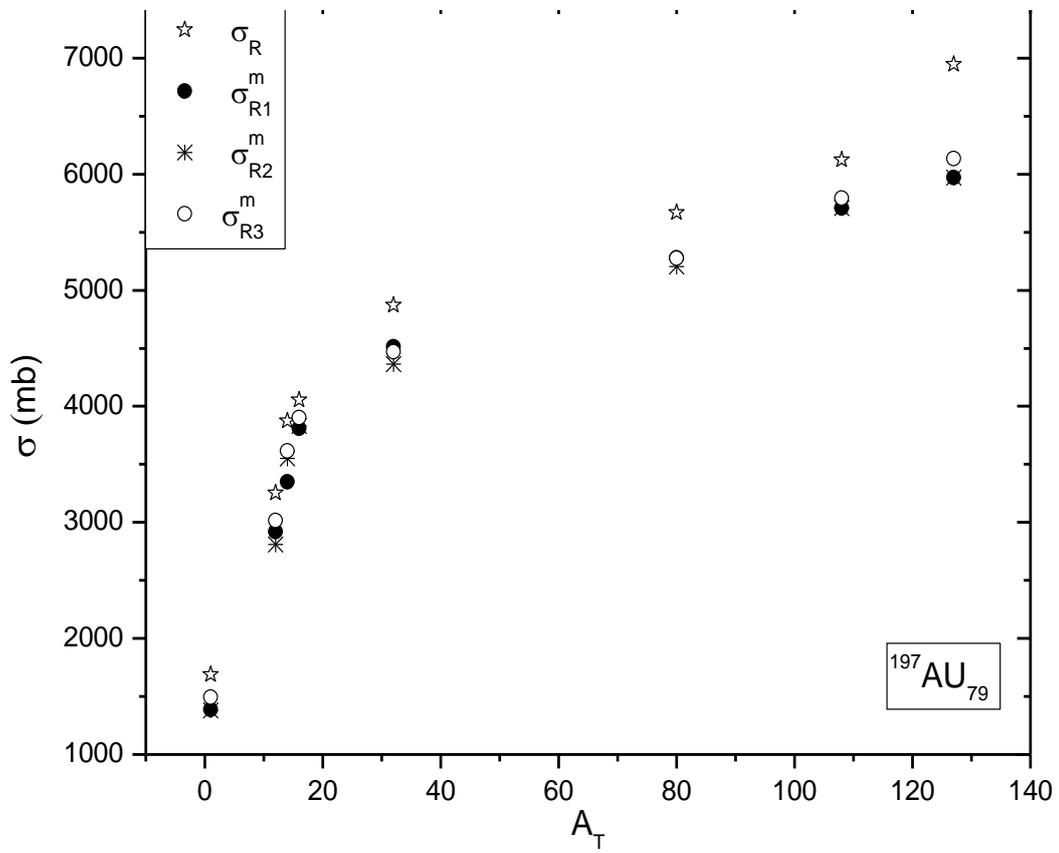

**(b)**

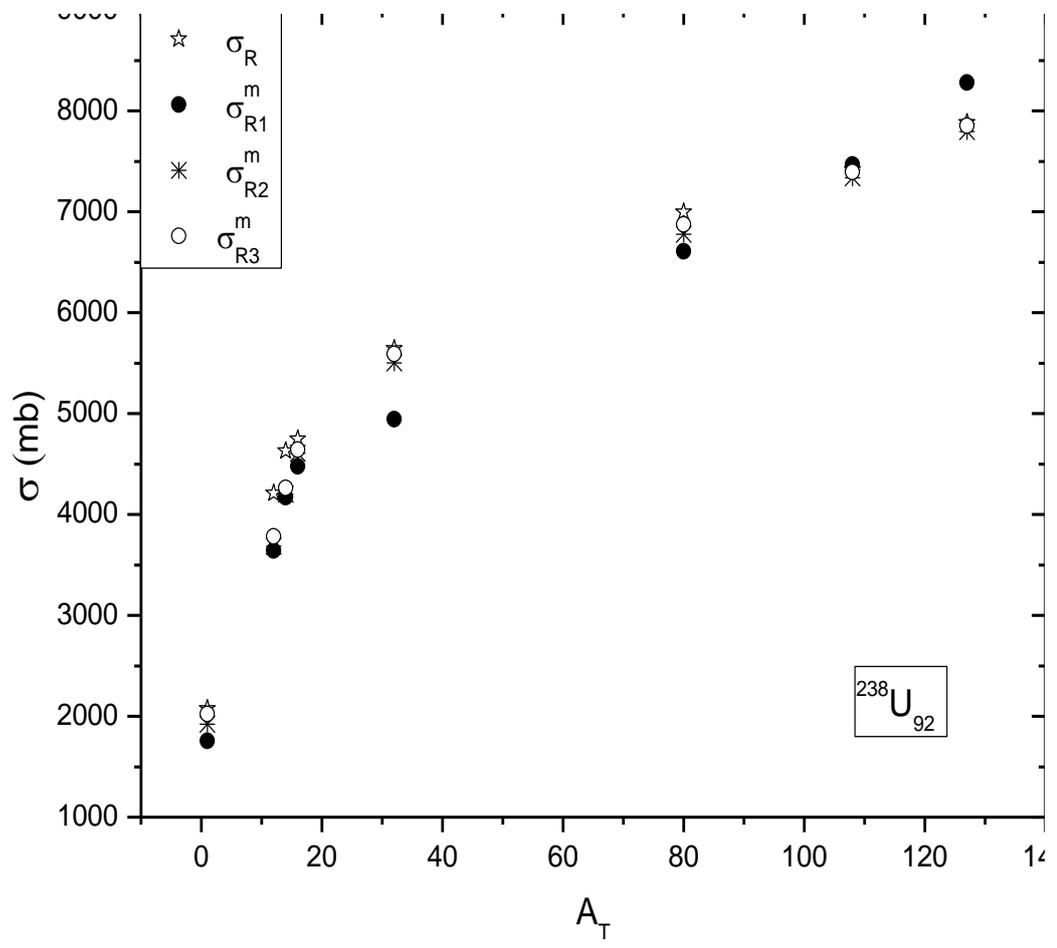

**(c)**

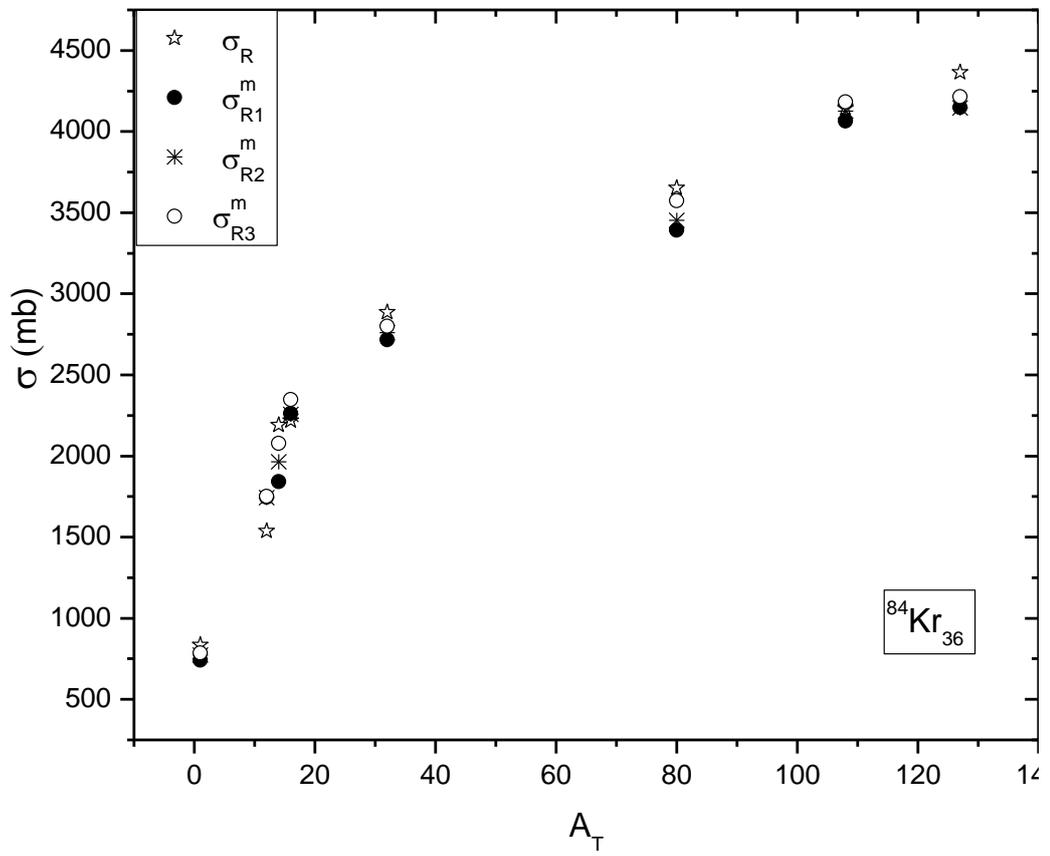

**(d)**

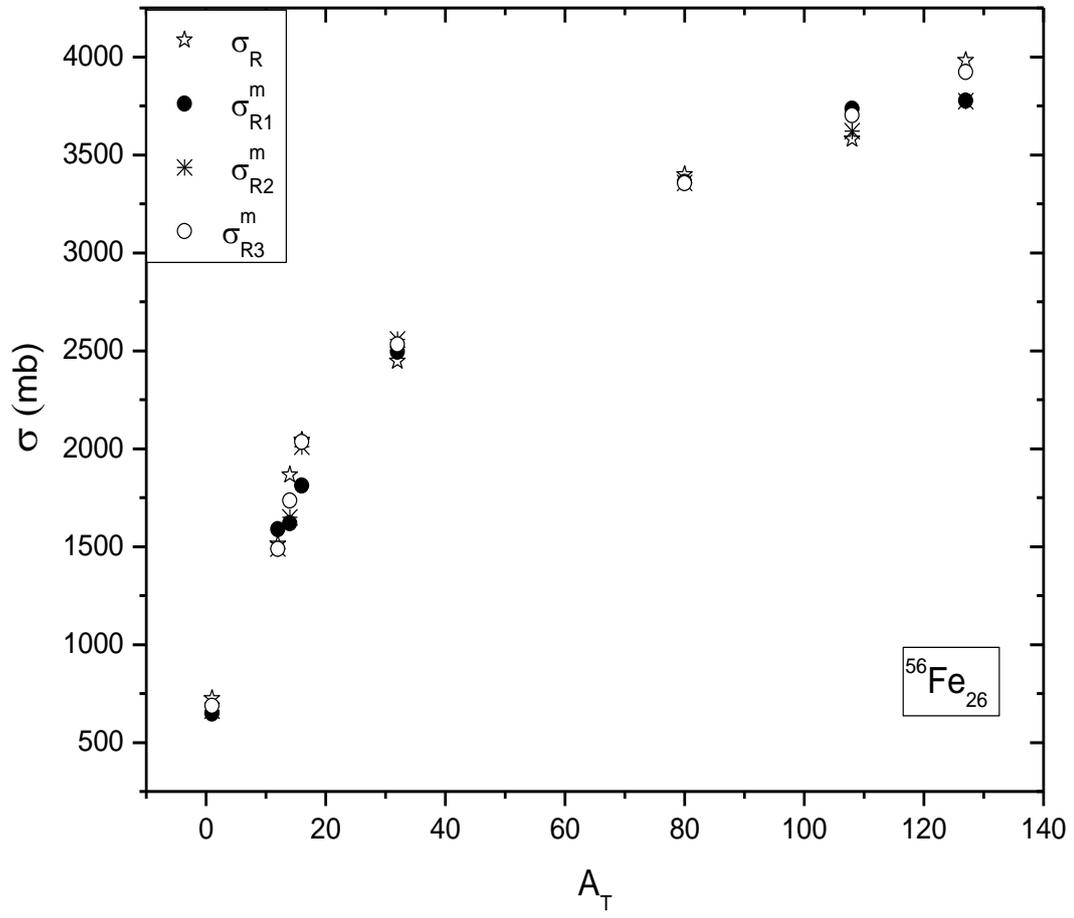

(e)

**Figure 2:** The total nuclear reactions cross-section of the projectile - emulsion with medium and without medium effect for the different projectiles as function of emulsion target mass ($A_T$) are shown from (a) to (e).

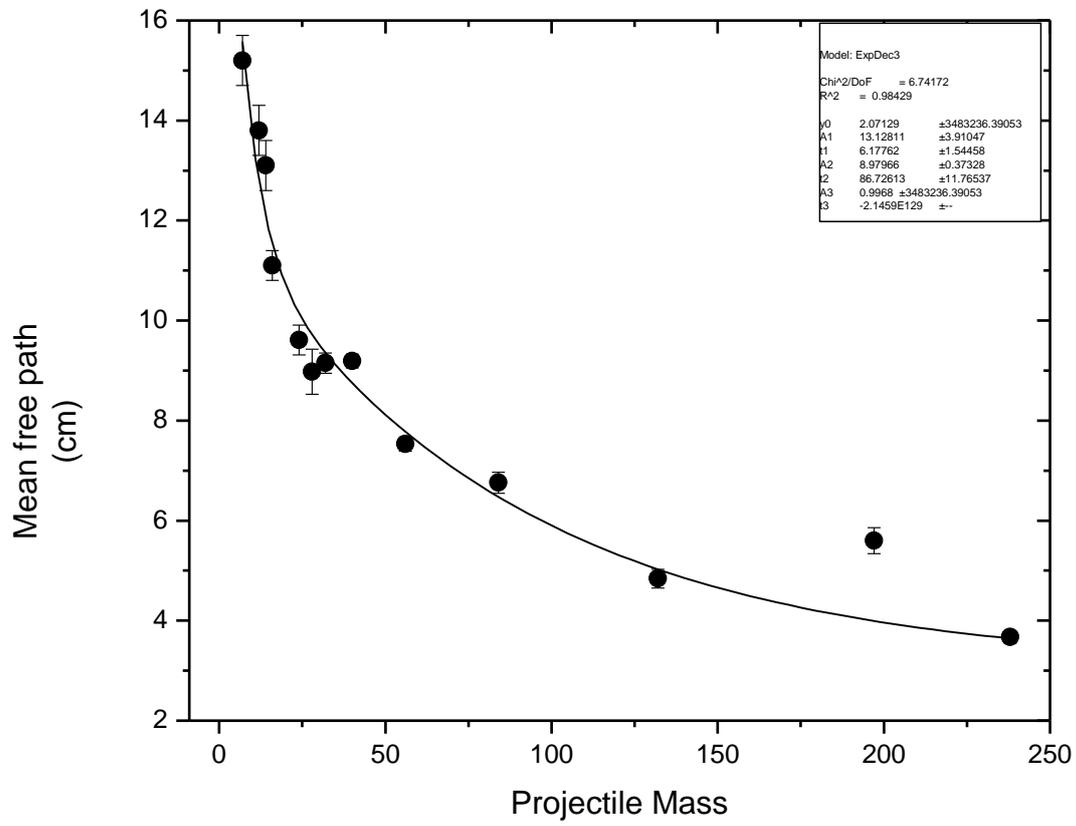

(a)

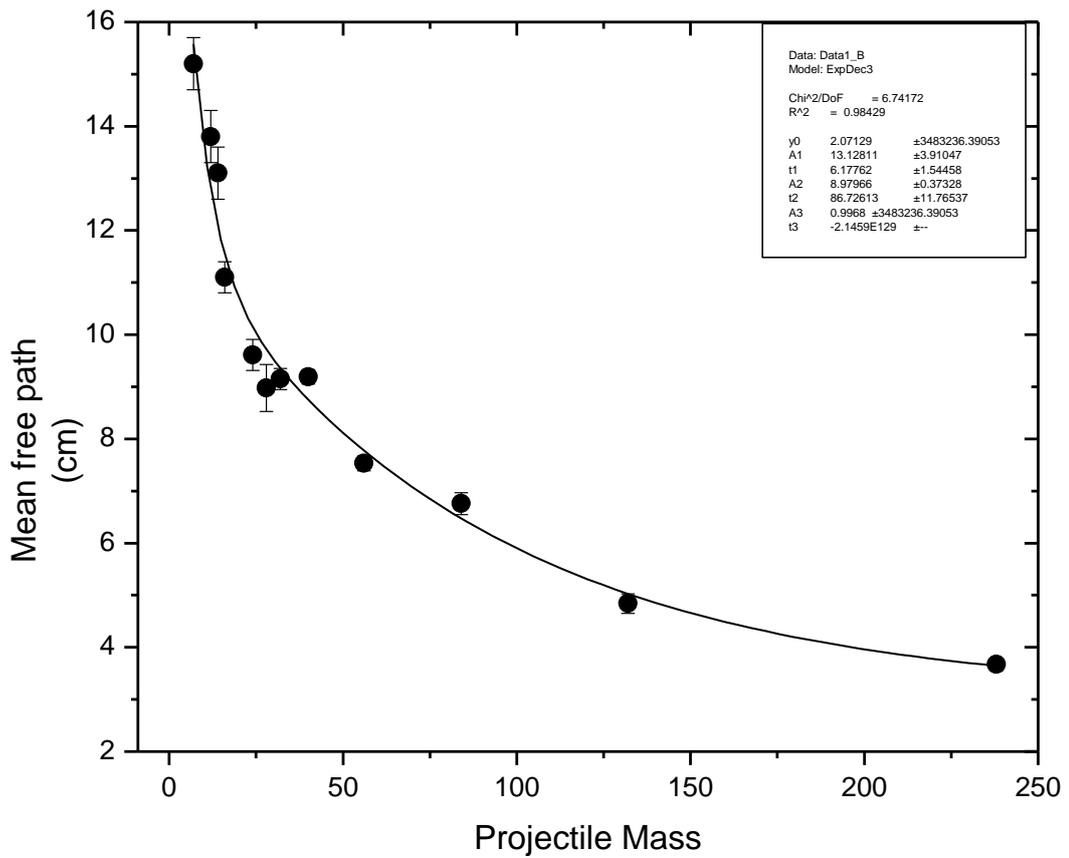

(b)

**Figure 3:** (a) The experimental mean free path of different projectiles as a function of a projectile mass number (b) without $^{197}Au_{79}$ projectile data.

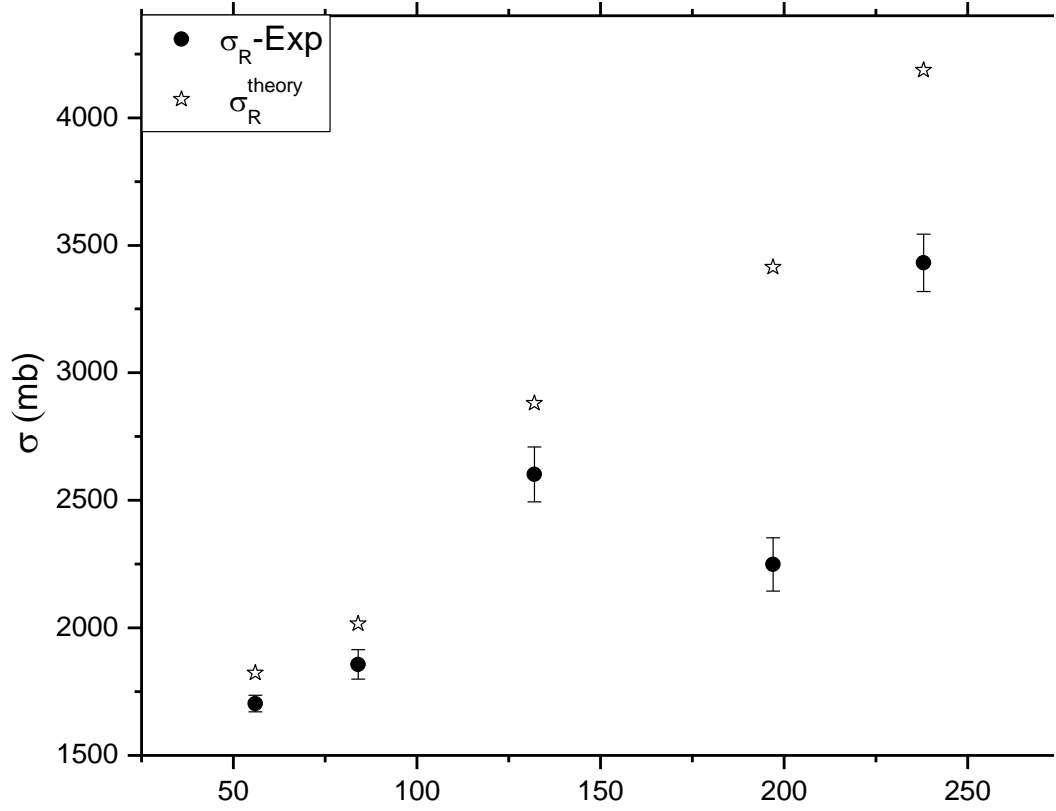

(a)

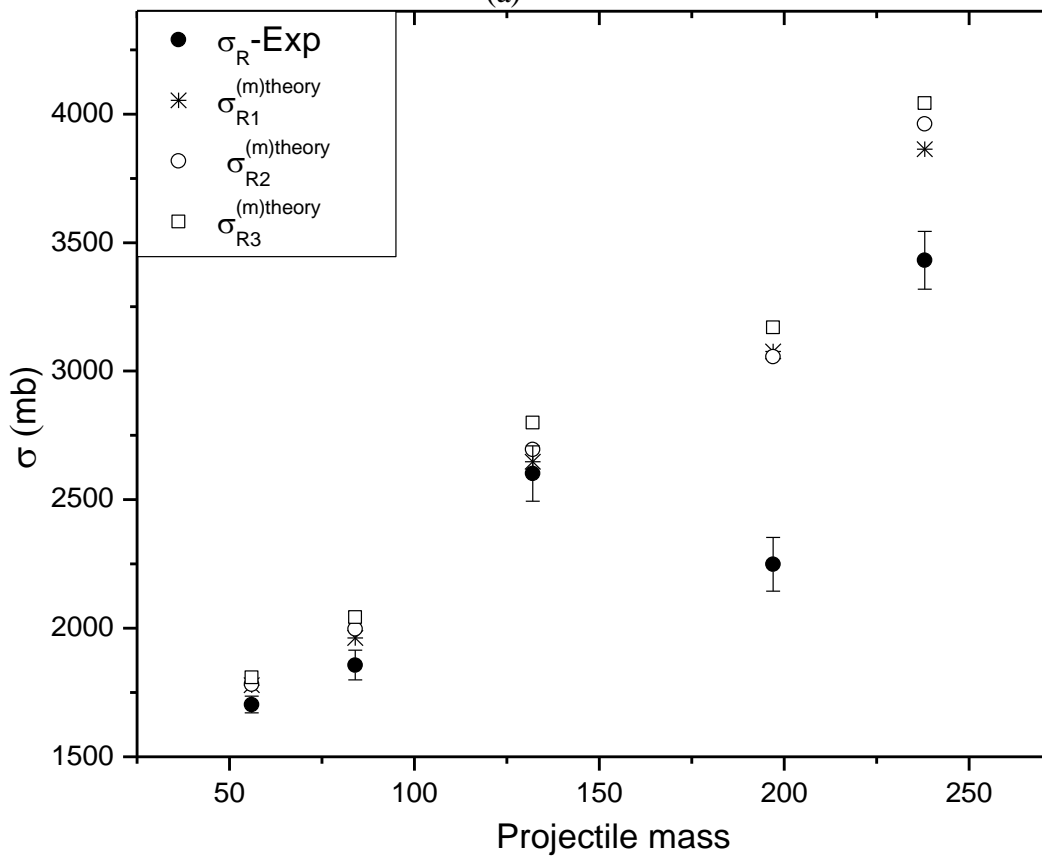

(b)

**Figure 4:** (a) The average value of total reaction cross-section considered without medium effect (b) The average value of total reaction cross-section considered with medium effect and both are corresponding to the experimental data for different projectiles $^{56}Fe_{26}$, $^{84}Kr_{36}$, $^{132}Xe_{54}$, $^{197}Au_{79}$ and $^{238}U_{92}$ at ~ 1 GeV / n.

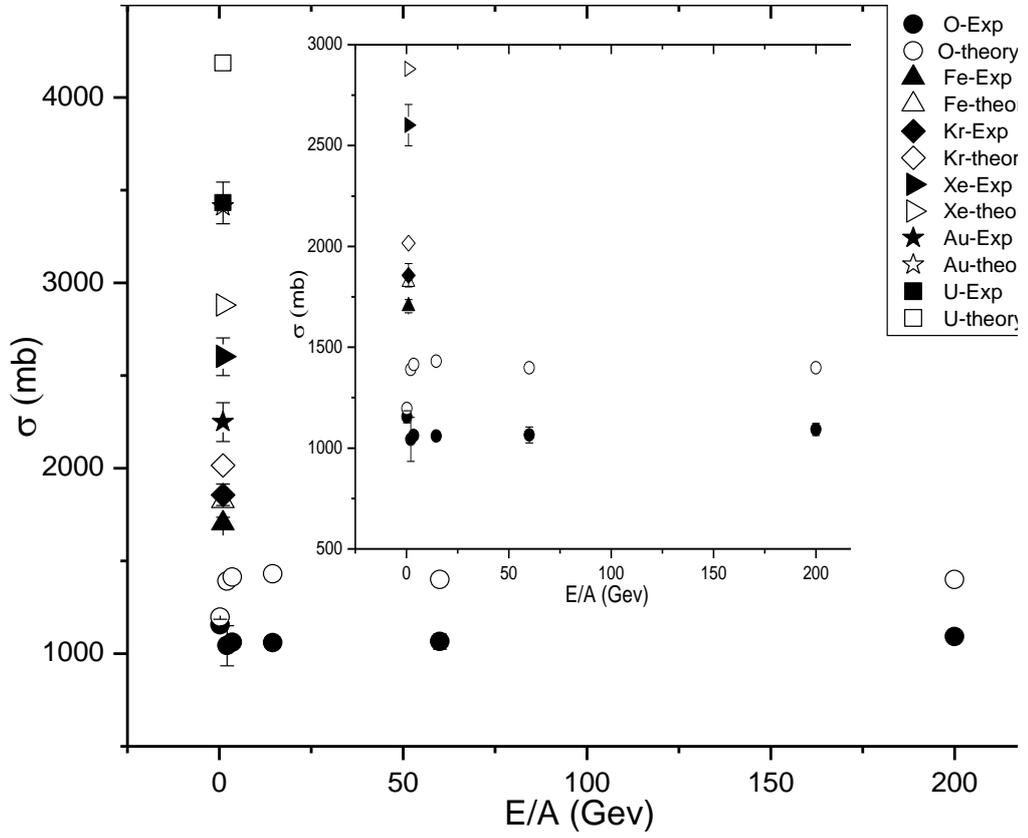

**Figure 5:** Energy dependence of the nuclear reaction cross-section for the $^{56}Fe_{26}$, $^{84}Kr_{36}$, $^{132}Xe_{54}$, $^{197}Au_{79}$, $^{238}U_{92}$ and $^{16}O_8$ projectiles. Inset plot is the zoomed one.

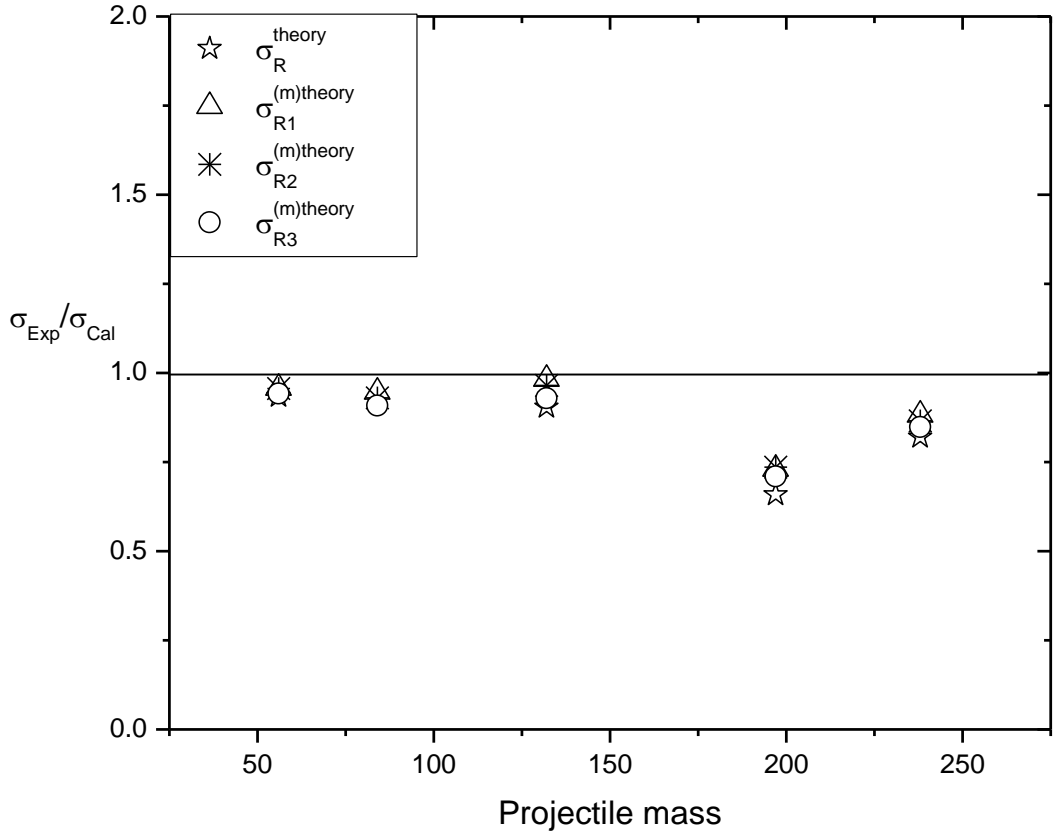

**Figure 6:** Ratio of $\sigma_{exp}/\sigma_{cal}$ as the function of projectile mass for $^{56}Fe_{26}$, $^{84}Kr_{36}$, $^{132}Xe_{54}$, $^{197}Au_{79}$ and $^{238}U_{92}$ at incident energies ~ 1 GeV/ n.

**Table 1:** The calculated total nuclear reactions cross-section without nuclear medium effect and with nuclear medium effect in case of Proton-Emulsion interactions for different projectiles at ~1 GeV/n is given below.

| Chemical Symbol | Mass Number | r m s radius (fm) | Ref. | No. of atoms ($10^{22}$ cm$^{-3}$) | Without medium effect $\sigma_R$ (mb) | With medium effect $\sigma^m_{R2}$ (mb) | $\sigma^{(ml)}_R$ (mb) | $\sigma^{m(ml)}_{R2}$ (mb) |
|---|---|---|---|---|---|---|---|---|
| \multicolumn{9}{c}{$^{56}$Fe-Em} |
| H  | 1   | 0.810 | [16] | 2.93  | 19.90   | 19.46   | 58.33   | 57.01   |
| C  | 12  | 2.442 | [6]  | 1.39  | 244.93  | 205.97  | 340.45  | 286.29  |
| N  | 14  | 2.580 | [30] | 0.37  | 273.54  | 217.5   | 101.21  | 80.47   |
| O  | 16  | 2.710 | [6]  | 1.06  | 299.33  | 256.28  | 317.29  | 271.65  |
| S  | 32  | 3.251 | [6]  | 0.004 | 467.13  | 457.29  | 1.86    | 1.82    |
| Br | 80  | 4.151 | [29] | 1.02  | 988.27  | 907.75  | 1008.03 | 925.90  |
| Ag | 108 | 4.542 | [6]  | 1.02  | 1140.55 | 1122.37 | 1163.36 | 1144.81 |
| I  | 127 | 4.749 | [29] | 0.003 | 1252.63 | 1178.33 | 3.75    | 3.53    |
| \multicolumn{9}{c}{$^{84}$Kr-Em} |
| H  | 1   | 0.810 |  | 2.93  | 23.51   | 19.48   | 68.88   | 57.07   |
| C  | 12  | 2.442 |  | 1.39  | 250.26  | 211.51  | 347.86  | 293.99  |
| N  | 14  | 2.580 |  | 0.37  | 280.23  | 243.85  | 103.68  | 90.22   |
| O  | 16  | 2.710 |  | 1.06  | 309.47  | 273.92  | 328.04  | 290.35  |
| S  | 32  | 3.251 |  | 0.004 | 501.83  | 429.16  | 2.00    | 1.71    |
| Br | 80  | 4.151 |  | 1.02  | 993.94  | 907.98  | 1013.82 | 926.13  |
| Ag | 108 | 4.542 |  | 1.02  | 1214.62 | 1089.16 | 1238.91 | 1110.94 |
| I  | 127 | 4.749 |  | 0.003 | 1322.8  | 1215.31 | 3.96    | 3.64    |
| \multicolumn{9}{c}{$^{132}$Xe-Em} |
| H  | 1   | 0.810 |  | 3.251 | 23.47   | 20.22   | 76.32   | 65.73   |
| C  | 12  | 2.442 |  | 1.39  | 251.07  | 212.86  | 348.99  | 295.87  |
| N  | 14  | 2.580 |  | 0.32  | 282.2   | 242.89  | 90.30   | 77.72   |
| O  | 16  | 2.710 |  | 0.94  | 314.26  | 278.93  | 295.40  | 262.19  |
| S  | 32  | 3.251 |  | 0.01  | 467.13  | 440.76  | 4.67    | 4.40    |
| Br | 80  | 4.151 |  | 1.01  | 951.98  | 910.44  | 961.49  | 919.54  |
| Ag | 108 | 4.542 |  | 1.02  | 1166.6  | 1126.6  | 1189.93 | 1149.13 |
| I  | 127 | 4.749 |  | 0.006 | 1253.04 | 1271.7  | 7.51    | 7.63    |
| \multicolumn{9}{c}{$^{197}$Au-Em} |
| H  | 1   | 0.810 |  | 3.251 | 23.45   | 20.48   | 76.24   | 66.58   |
| C  | 12  | 2.442 |  | 1.39  | 252.03  | 217.23  | 350.32  | 301.94  |
| N  | 14  | 2.580 |  | 0.32  | 273.54  | 251.86  | 87.53   | 80.59   |
| O  | 16  | 2.710 |  | 0.94  | 290.10  | 280.03  | 272.69  | 263.22  |
| S  | 32  | 3.251 |  | 0.01  | 458.10  | 457.29  | 4.58    | 4.57    |
| Br | 80  | 4.151 |  | 1.01  | 988.66  | 867.5   | 998.55  | 876.17  |
| Ag | 108 | 4.542 |  | 1.02  | 1198.87 | 1115.45 | 1222.84 | 1137.75 |
| I  | 127 | 4.749 |  | 0.006 | 1323.06 | 1275.43 | 7.93    | 7.65    |
| \multicolumn{9}{c}{$^{238}$U-Em} |
| H  | 1   | 0.810 |  | 3.251 | 23.91   | 20.63   | 77.73   | 67.06   |
| C  | 12  | 2.442 |  | 1.39  | 250.26  | 222.11  | 347.86  | 308.73  |
| N  | 14  | 2.580 |  | 0.32  | 284.80  | 254.23  | 91.13   | 81.35   |
| O  | 16  | 2.710 |  | 0.94  | 309.47  | 283.87  | 290.90  | 266.83  |
| S  | 32  | 3.251 |  | 0.01  | 532.65  | 468.17  | 5.32    | 4.68    |
| Br | 80  | 4.151 |  | 1.01  | 994.21  | 930.91  | 1004.15 | 940.21  |
| Ag | 108 | 4.542 |  | 1.02  | 1224.28 | 1155.75 | 1248.76 | 1178.86 |
| I  | 127 | 4.749 |  | 0.006 | 1353.01 | 1301.13 | 8.11    | 7.80    |

**Table 2:** The calculated total nuclear reactions cross-section without medium effect and with nuclear medium effect in case of Projectile-Emulsion interactions for different projectiles at ~1 GeV / n is given below.

| Chemical Symbol | Mass Number | rms radius (fm) | Ref. | No. of atoms ($10^{22}$ cm$^{-3}$) | Without medium effect $\sigma_R$ (mb) | With medium effect $\sigma_{R2}^m$ (mb) | $\sigma_R^{(ml)}$ (mb) | $\sigma_{R2}^{m(ml)}$ (mb) |
|---|---|---|---|---|---|---|---|---|
| $^{56}$Fe-Em | | | | | | | | |
| H  | 1   | 0.810 | [16] | 2.93  | 724.14  | 661.83  | 2121.75 | 1939.16 |
| C  | 12  | 2.442 | [6]  | 1.39  | 1511.26 | 1490.72 | 2100.65 | 2072.10 |
| N  | 14  | 2.580 | [30] | 0.37  | 1865.27 | 1648.82 | 690.14  | 610.06  |
| O  | 16  | 2.710 | [6]  | 1.06  | 2041.85 | 2010.7  | 2164.36 | 2131.34 |
| S  | 32  | 3.251 | [6]  | 0.004 | 2445.98 | 2558.86 | 9.78    | 10.23   |
| Br | 80  | 4.151 | [29] | 1.02  | 3398.53 | 3357.72 | 3466.50 | 3424.87 |
| Ag | 108 | 4.542 | [6]  | 1.02  | 3580.34 | 3620.3  | 3651.94 | 3692.70 |
| I  | 127 | 4.749 | [29] | 0.003 | 3981.55 | 3774.63 | 11.94   | 11.32   |
| $^{84}$Kr-Em | | | | | | | | |
| H  | 1   | 0.810 | | 2.93  | 834.48  | 773.568 | 2445.02 | 2266.55 |
| C  | 12  | 2.442 | | 1.39  | 1537.68 | 1743.98 | 2137.37 | 2424.13 |
| N  | 14  | 2.580 | | 0.37  | 2189.92 | 1965.44 | 810.27  | 727.21  |
| O  | 16  | 2.710 | | 1.06  | 2218.58 | 2257.47 | 2351.69 | 2392.91 |
| S  | 32  | 3.251 | | 0.004 | 2885.14 | 2759.37 | 11.54   | 11.03   |
| Br | 80  | 4.151 | | 1.02  | 3651.26 | 3452.15 | 3724.28 | 3521.19 |
| Ag | 108 | 4.542 | | 1.02  | 4142.35 | 4127.27 | 4225.19 | 4209.81 |
| I  | 127 | 4.749 | | 0.003 | 4365.48 | 4146.67 | 13.09   | 12.44   |
| $^{132}$Xe-Em | | | | | | | | |
| H  | 1   | 0.810 | | 3.251 | 1378.32 | 1116.52 | 4480.91 | 3629.80 |
| C  | 12  | 2.442 | | 1.39  | 2830.91 | 2429.46 | 3934.96 | 3376.94 |
| N  | 14  | 2.580 | | 0.32  | 3143.44 | 2500.9  | 1005.90 | 800.28  |
| O  | 16  | 2.710 | | 0.94  | 3326.98 | 2955.95 | 3127.36 | 2778.59 |
| S  | 32  | 3.251 | | 0.01  | 3690.55 | 3818.11 | 36.90   | 38.18   |
| Br | 80  | 4.151 | | 1.01  | 4841.47 | 5052.64 | 4889.88 | 5103.16 |
| Ag | 108 | 4.542 | | 1.02  | 5269.82 | 5539.67 | 5375.21 | 5650.46 |
| I  | 127 | 4.749 | | 0.006 | 5675.24 | 5939.52 | 34.05   | 35.63   |
| $^{197}$Au-Em | | | | | | | | |
| H  | 1   | 0.810 | | 3.251 | 1687.89 | 1380.52 | 5487.33 | 4488.07 |
| C  | 12  | 2.442 | | 1.39  | 3254.06 | 2810.75 | 4523.14 | 3906.94 |
| N  | 14  | 2.580 | | 0.32  | 3874.18 | 3552.23 | 1239.73 | 1136.71 |
| O  | 16  | 2.710 | | 0.94  | 4055.61 | 3831.86 | 3812.27 | 3601.94 |
| S  | 32  | 3.251 | | 0.01  | 4871.57 | 4365.15 | 48.71   | 43.65   |
| Br | 80  | 4.151 | | 1.01  | 5668.62 | 5203.02 | 5725.30 | 5255.05 |
| Ag | 108 | 4.542 | | 1.02  | 6125.13 | 5709.78 | 6247.63 | 5823.97 |
| I  | 127 | 4.749 | | 0.006 | 6945.37 | 5969.96 | 41.67   | 35.81   |
| $^{238}$U-Em | | | | | | | | |
| H  | 1   | 0.810 | | 3.251 | 2075.47 | 1923.58 | 6747.35 | 6253.55 |
| C  | 12  | 2.442 | | 1.39  | 4208.42 | 3687.58 | 5849.70 | 5125.73 |
| N  | 14  | 2.580 | | 0.32  | 4629.96 | 4199.16 | 1481.58 | 1343.73 |
| O  | 16  | 2.710 | | 0.94  | 4747.73 | 4607.68 | 4462.86 | 4331.21 |
| S  | 32  | 3.251 | | 0.01  | 5641.93 | 5500.35 | 56.41   | 55.00   |
| Br | 80  | 4.151 | | 1.01  | 6994.17 | 6777.82 | 7064.11 | 6845.59 |
| Ag | 108 | 4.542 | | 1.02  | 7413.16 | 7334.71 | 7561.42 | 7481.40 |
| I  | 127 | 4.749 | | 0.006 | 7883.63 | 7795.32 | 47.30   | 46.77   |

**Table 3:** Amount of nuclear matter involved in the proton-emulsion interaction for different projectiles with nuclear emulsion target at incident energies ~ 1 GeV/n is estimated. These calculated values are averaged over the impact parameter.

| Chemical Symbol | Mass Number | rms radius (fm) | Without medium effect $\sigma_R$ (mb) | With in-medium effect $\sigma_{R2}^m$ (mb) | Without - medium effect | | | With in-medium effect | | |
|---|---|---|---|---|---|---|---|---|---|---|
| | | | | | $P_{proj}$ | $P_{targ}$ | $B_C$ | $P_{proj}$ | $P_{targ}$ | $B_C$ |
| $^{56}$Fe-Em | | | | | | | | | | |
| H | 1 | 0.810 | 19.90 | 19.46 | 1.91 | 1.91 | 1.62 | 1.96 | 1.96 | 1.65 |
| C | 12 | 2.442 | 244.93 | 205.97 | 1 | 1.87 | 1.58 | 1.10 | 2.22 | 1.88 |
| N | 14 | 2.580 | 273.54 | 217.5 | 1 | 1.95 | 1.65 | 1.17 | 2.45 | 2.07 |
| O | 16 | 2.710 | 299.33 | 256.28 | 1 | 2.04 | 1.72 | 1.09 | 2.38 | 2.01 |
| S | 32 | 3.251 | 467.13 | 457.29 | 1 | 2.61 | 2.21 | 1 | 2.67 | 2.26 |
| Br | 80 | 4.151 | 988.27 | 907.75 | 1 | 3.08 | 2.61 | 1 | 3.36 | 2.84 |
| Ag | 108 | 4.542 | 1140.55 | 1122.37 | 1 | 3.61 | 3.05 | 1 | 3.67 | 3.10 |
| I | 127 | 4.749 | 1252.63 | 1178.33 | 1 | 3.86 | 3.27 | 1.05 | 4.11 | 3.48 |
| $^{84}$Kr-Em | | | | | | | | | | |
| H | 1 | 0.810 | 23.51 | 19.48 | 1.62 | 1.62 | 1.37 | 1.95 | 1.95 | 1.65 |
| C | 12 | 2.442 | 250.26 | 211.51 | 1 | 1.83 | 1.54 | 1.07 | 2.16 | 1.83 |
| N | 14 | 2.580 | 280.23 | 243.85 | 1 | 1.90 | 1.61 | 1.04 | 2.19 | 1.85 |
| O | 16 | 2.710 | 309.47 | 273.92 | 1 | 1.97 | 1.66 | 1.02 | 2.22 | 1.88 |
| S | 32 | 3.251 | 501.83 | 429.16 | 1 | 2.43 | 2.05 | 1.07 | 2.84 | 2.40 |
| Br | 80 | 4.151 | 993.94 | 907.98 | 1 | 3.07 | 2.59 | 1 | 3.36 | 2.84 |
| Ag | 108 | 4.542 | 1214.62 | 1089.16 | 1 | 3.39 | 2.87 | 1.01 | 3.78 | 3.20 |
| I | 127 | 4.749 | 1322.8 | 1215.31 | 1 | 3.6 | 3.10 | 1.02 | 3.98 | 3.37 |
| $^{131}$Xe-Em | | | | | | | | | | |
| H | 1 | 0.810 | 23.47 | 20.22 | 1.62 | 1.62 | 1.37 | 1.88 | 1.88 | 1.59 |
| C | 12 | 2.442 | 251.07 | 212.86 | 1 | 1.82 | 1.54 | 1.07 | 2.15 | 1.82 |
| N | 14 | 2.580 | 282.2 | 242.89 | 1 | 1.89 | 1.60 | 1.04 | 2.20 | 1.86 |
| O | 16 | 2.710 | 314.26 | 278.93 | 1 | 1.94 | 1.64 | 1 | 2.18 | 1.85 |
| S | 32 | 3.251 | 467.13 | 440.76 | 1 | 2.61 | 2.21 | 1.04 | 2.77 | 2.34 |
| Br | 80 | 4.151 | 951.98 | 910.44 | 1 | 3.20 | 2.71 | 1 | 3.35 | 2.83 |
| Ag | 108 | 4.542 | 1166.6 | 1126.6 | 1 | 3.53 | 2.99 | 1 | 3.65 | 3.09 |
| I | 127 | 4.749 | 1253.04 | 1271.7 | 1 | 3.86 | 3.27 | 1 | 3.81 | 3.22 |
| $^{197}$Au-Em | | | | | | | | | | |
| H | 1 | 0.810 | 23.45 | 20.48 | 1.62 | 1.62 | 1.37 | 1.86 | 1.86 | 1.57 |
| C | 12 | 2.442 | 252.03 | 217.23 | 1 | 1.81 | 1.53 | 1.04 | 2.10 | 1.78 |
| N | 14 | 2.580 | 273.54 | 251.86 | 1 | 1.95 | 1.65 | 1.01 | 2.12 | 1.79 |
| O | 16 | 2.710 | 290.10 | 280.03 | 1 | 2.10 | 1.78 | 1 | 2.18 | 1.84 |
| S | 32 | 3.251 | 458.10 | 457.29 | 1 | 2.66 | 2.25 | 1 | 2.67 | 2.26 |
| Br | 80 | 4.151 | 988.66 | 867.5 | 1 | 3.08 | 2.61 | 1.02 | 3.52 | 2.97 |
| Ag | 108 | 4.542 | 1198.87 | 1115.45 | 1 | 3.43 | 2.90 | 1 | 3.69 | 3.12 |
| I | 127 | 4.749 | 1323.06 | 1275.43 | 1 | 3.66 | 3.10 | 1 | 3.80 | 3.21 |
| $^{238}$U-Em | | | | | | | | | | |
| H | 1 | 0.810 | 23.91 | 20.63 | 1.59 | 1.59 | 1.35 | 1.85 | 1.85 | 1.56 |
| C | 12 | 2.442 | 250.26 | 222.11 | 1 | 1.83 | 1.54 | 1.02 | 2.06 | 1.74 |
| N | 14 | 2.580 | 284.80 | 254.23 | 1 | 1.87 | 1.58 | 1 | 2.10 | 1.77 |
| O | 16 | 2.710 | 309.47 | 283.87 | 1 | 1.97 | 1.66 | 1 | 2.15 | 1.82 |
| S | 32 | 3.251 | 532.65 | 468.17 | 1 | 2.29 | 1.94 | 1 | 2.60 | 2.20 |
| Br | 80 | 4.151 | 994.21 | 930.91 | 1 | 3.07 | 2.59 | 1 | 3.28 | 2.77 |
| Ag | 108 | 4.542 | 1224.28 | 1155.75 | 1 | 3.36 | 2.84 | 1 | 3.56 | 3.01 |
| I | 127 | 4.749 | 1353.01 | 1301.13 | 1 | 3.58 | 3.03 | 1 | 3.72 | 3.15 |

**Table 4:** The amount of nuclear matter involved in the projectile-emulsion interaction for different projectiles with nuclear emulsion target at incident energies ~ 1 GeV / n is estimated. These calculated values are averaged over the impact parameter.

| Chemical Symbol | Mass Number | r m s radius (fm) | Without medium effect $\sigma_R$ (mb) | With medium effect $\sigma_{R2}^m$ (mb) | Without medium effect | | | With medium effect | | |
|---|---|---|---|---|---|---|---|---|---|---|
| | | | | | $P_{proj}$ | $P_{targ}$ | $B_C$ | $P_{proj}$ | $P_{targ}$ | $B_C$ |
| $^{56}$Fe-Em | | | | | | | | | | |
| H  | 1   | 0.810 | 724.14  | 661.83  | 2.95  | 1     | 2.49  | 3.22  | 1.04  | 2.73  |
| C  | 12  | 2.442 | 1511.26 | 1490.72 | 8.44  | 5.47  | 14.36 | 8.55  | 5.55  | 14.56 |
| N  | 14  | 2.580 | 1865.27 | 1648.82 | 7.64  | 5.17  | 13.57 | 8.64  | 5.85  | 15.35 |
| O  | 16  | 2.710 | 2041.85 | 2010.7  | 7.68  | 5.40  | 14.17 | 7.80  | 5.48  | 14.39 |
| S  | 32  | 3.251 | 2445.98 | 2558.86 | 10.55 | 9.02  | 23.66 | 10.09 | 8.62  | 22.62 |
| Br | 80  | 4.151 | 3398.53 | 3357.72 | 14.68 | 16.23 | 42.57 | 14.86 | 16.43 | 43.09 |
| Ag | 108 | 4.542 | 3580.34 | 3620.3  | 17.29 | 20.80 | 54.56 | 17.10 | 20.57 | 53.95 |
| I  | 127 | 4.749 | 3981.55 | 3774.63 | 17.47 | 21.99 | 57.69 | 18.43 | 23.20 | 60.85 |
| $^{84}$Kr-Em | | | | | | | | | | |
| H  | 1   | 0.810 | 834.48  | 773.568 | 3.84  | 1.10  | 3.25  | 4.14  | 1.19  | 3.50  |
| C  | 12  | 2.442 | 1537.68 | 1743.98 | 12.44 | 7.20  | 21.17 | 10.97 | 6.35  | 18.66 |
| N  | 14  | 2.580 | 2189.92 | 1965.44 | 9.76  | 5.90  | 17.34 | 10.87 | 6.57  | 19.32 |
| O  | 16  | 2.710 | 2218.58 | 2257.47 | 10.60 | 6.65  | 19.56 | 10.42 | 6.54  | 19.23 |
| S  | 32  | 3.251 | 2885.14 | 2759.37 | 13.42 | 10.23 | 30.09 | 14.04 | 10.70 | 31.46 |
| Br | 80  | 4.151 | 3651.26 | 3452.15 | 20.50 | 20.22 | 59.44 | 21.68 | 21.39 | 62.87 |
| Ag | 108 | 4.542 | 4142.35 | 4127.27 | 22.42 | 24.06 | 70.73 | 22.50 | 24.15 | 70.99 |
| I  | 127 | 4.749 | 4365.48 | 4146.67 | 23.91 | 26.85 | 78.93 | 25.17 | 28.27 | 83.09 |
| $^{131}$Xe-Em | | | | | | | | | | |
| H  | 1   | 0.810 | 1378.32 | 1116.52 | 3.65  | 1     | 3.09  | 4.51  | 1.144 | 3.81  |
| C  | 12  | 2.442 | 2830.91 | 2429.46 | 10.62 | 5.41  | 18.07 | 12.37 | 6.31  | 21.05 |
| N  | 14  | 2.580 | 3143.44 | 2500.9  | 10.68 | 5.69  | 18.98 | 13.43 | 7.15  | 23.86 |
| O  | 16  | 2.710 | 3326.98 | 2955.95 | 11.11 | 6.14  | 20.50 | 12.51 | 6.91  | 23.07 |
| S  | 32  | 3.251 | 3690.55 | 3818.11 | 16.49 | 11.07 | 36.96 | 15.94 | 10.70 | 35.73 |
| Br | 80  | 4.151 | 4841.47 | 5052.64 | 24.30 | 21.11 | 70.45 | 23.28 | 20.22 | 67.50 |
| Ag | 108 | 4.542 | 5269.82 | 5539.67 | 27.70 | 26.18 | 87.37 | 26.35 | 24.90 | 83.12 |
| I  | 127 | 4.749 | 5675.24 | 5939.52 | 28.90 | 28.59 | 95.41 | 27.61 | 27.31 | 91.16 |
| $^{197}$Au-Em | | | | | | | | | | |
| H  | 1   | 0.810 | 1687.89 | 1380.52 | 4.45  | 1     | 3.76   | 5.44  | 1.23  | 4.60   |
| C  | 12  | 2.442 | 3254.06 | 2810.75 | 13.79 | 6.28  | 23.46  | 15.96 | 7.27  | 27.16  |
| N  | 14  | 2.580 | 3874.18 | 3552.23 | 12.94 | 6.15  | 22.99  | 14.11 | 6.71  | 25.07  |
| O  | 16  | 2.710 | 4055.61 | 3831.86 | 13.61 | 6.72  | 25.10  | 14.40 | 7.11  | 26.56  |
| S  | 32  | 3.251 | 4871.57 | 4365.15 | 18.65 | 11.19 | 41.79  | 20.81 | 12.49 | 46.64  |
| Br | 80  | 4.151 | 5668.62 | 5203.02 | 30.97 | 24.04 | 89.80  | 33.74 | 26.19 | 97.83  |
| Ag | 108 | 4.542 | 6125.13 | 5709.78 | 35.57 | 30.04 | 112.19 | 38.15 | 32.22 | 120.35 |
| I  | 127 | 4.749 | 6945.37 | 5969.96 | 35.24 | 31.15 | 116.35 | 41    | 36.24 | 135.36 |
| $^{238}$U-Em | | | | | | | | | | |
| H  | 1   | 0.810 | 2075.47 | 1923.58 | 4.37  | 1     | 3.70   | 4.72  | 1.01  | 3.99   |
| C  | 12  | 2.442 | 4208.42 | 3687.58 | 12.88 | 5.56  | 21.92  | 14.70 | 6.35  | 25.01  |
| N  | 14  | 2.580 | 4629.96 | 4199.16 | 13.08 | 5.90  | 23.24  | 14.42 | 6.50  | 25.62  |
| O  | 16  | 2.710 | 4747.73 | 4607.68 | 14.04 | 6.57  | 25.90  | 14.47 | 6.77  | 26.69  |
| S  | 32  | 3.251 | 5641.93 | 5500.35 | 19.45 | 11.07 | 43.60  | 19.95 | 11.35 | 44.72  |
| Br | 80  | 4.151 | 6994.17 | 6777.82 | 30.33 | 22.32 | 87.92  | 31.29 | 23.03 | 90.73  |
| Ag | 108 | 4.542 | 7413.16 | 7334.71 | 35.50 | 28.43 | 111.99 | 35.88 | 28.74 | 113.19 |
| I  | 127 | 4.749 | 7883.63 | 7795.32 | 37.51 | 31.44 | 123.83 | 37.94 | 31.80 | 125.24 |

**Table 5:** The calculated average values of shower particles multiplicities in the framework of CGCM with and with nuclear medium effect approaches compared with corresponding experimental values at ~ 1 GeV / n.

| Reaction Systems | $<n_s>^{Exp}$ | Ref. | Calculation with nuclear medium effect | | | Calculation without nuclear medium effect | | |
|---|---|---|---|---|---|---|---|---|
| | | | $<n_s>$ $(P_{prog} + P_{targ})$ | $<n_s>_{B_C}$ | $<n_s>^{theory}$ | $<n_s>$ $(P_{prog} + P_{targ})$ | $<n_s>_{B_C}$ | $<n_s>^{theory}$ |
| $^{56}$Fe-Em | - | - | 21.47 | 27.88 | 10.26 | 21.93 | 28.44 | 9.64 |
| $^{84}$Kr-Em | 13.14±0.39 | [9] | 27.39 | 37.56 | 14.29 | 28.11 | 38.64 | 13.05 |
| $^{132}$Xe-Em | 17.40±0.70 | [38] | 29.82 | 43.85 | 15.90 | 30.08 | 43.66 | 14.83 |
| $^{197}$Au-Em | 16.43±3.43 | [34] | 35.23 | 54.43 | 19.59 | 39.13 | 60.45 | 20.24 |
| $^{238}$U-Em | - | - | 34.93 | 55.26 | 20.48 | 36.11 | 56.90 | 19.49 |